\documentclass[12pt]{iopart}

\usepackage{iopams}
\usepackage{graphicx}
\usepackage{dcolumn}
\usepackage{bm}
\usepackage{rotating}
\usepackage{mathbbol}

\begin{document}

\title[Schumpeterian Dynamics]{Schumpeterian economic dynamics  as a quantifiable minimum model of evolution}

\author{Stefan Thurner$^{1,2,*}$, Peter Klimek$^1$, Rudolf Hanel$^1$}

\address{$^1$Complex Systems Research Group, Medical University of Vienna, W\"ahringer G\"urtel 18-20, A-1090, Austria\\ 
$^2$Santa Fe Institute, 1399 Hyde Park Road, Santa Fe, NM 87501, USA}

\ead{$^*$thurner@univie.ac.at}

\begin{abstract}
We propose a simple quantitative model of Schumpeterian economic dynamics. 
New goods and services are endogenously produced through combinations of existing goods. 
As soon as new goods enter the market they may compete against already existing goods, in other words
new products  can have 
destructive effects on existing goods. As a result of this competition mechanism 
existing goods may be driven out from the market -- often causing cascades of secondary defects (Schumpeterian 
gales of destruction). 
The model leads to a generic dynamics characterized by phases of relative economic stability followed by phases 
of massive restructuring of markets -- which could be interpreted as Schumpeterian  business `cycles'. 
Model timeseries of product diversity and productivity  reproduce several stylized facts of economics timeseries on long 
timescales such as GDP or business failures, including non-Gaussian fat tailed distributions, volatility clustering etc. 
The model is phrased in an open, non-equilibrium setup which can be understood as a self organized critical system. 
Its diversity dynamics can be understood by the time-varying topology of the active production networks. \\

\end{abstract}

\maketitle

\section{Introduction}
The essence of the work of Joseph Schumpeter  is to understand economic development and growth as an evolutionary process, out of equilibrium,   
driven by the appearance and disappearance of goods and services. Goods and services appear and disappear  endogenously as a result of 
technological progress and innovation, which is driven by 
market participants (firms and consumers) maximizing their respective profit or utility functions. 
Trying to understand capitalist economy without these concepts is ``... like Hamlet without the Danish prince'' \cite{schumpeter1942} 
\footnote{From \cite{schumpeter1942} chapter 8:  
``The essential point to grasp is that in dealing with capitalism we are dealing with an evolutionary process [...] 
The fundamental impulse that sets and keeps the capitalist engine in motion comes from the new consumer goods, the new methods of production,
or transportation, the new forms of industrial organization that capitalist enterprise creates [...] In the case of retail trade the competition that matters arises not from additional shops of the same type, but from the department store, the chain store, the mail-order house and the super market, which are bound to destroy those pyramids sooner or later. Now a theoretical construction which neglects this essential element of the case neglects all that is most typically capitalist about it; even if correct in logic as well as in fact, it is like Hamlet without the Danish prince.''}.

Schumpeterian growth is based on the endogenous introduction of new goods, products, processes or services 
and is governed by the process of {\em creative destruction} \cite{schumpeter1939,schumpeter1942}. Creative 
destruction means that the appearance of a new good (through a successful innovation) can have devastating 
effects on seemingly well established goods, eventually driving them out of business. Examples include the collapse of 
horse carriage industry with the innovation of the combustion engine, or the disappearance of Polaroid cameras with the 
invention of the digital camera. 
The process of creative  destruction is sometimes referred to as {\em gales of destruction}, pointing to the fact that 
economic consequences of innovation can be severe and massive. 
Excluding  innovations  leads to a stationary state which is described by Walrasian equilibrium \cite{schumpeter1911}.
Entrepreneurs, by transforming ideas into innovations,  disturb this equilibrium and cause economic development. 
With this view Schumpeter aimed at a  qualitative understanding of empirical economic facts such as business cycles, 
or fluctuations \cite{schumpeter1939}. 

Surprisingly,  current main stream economics, general equilibrium theory in particular, systematically 
focuses on situations where none of the above elements are present. By excluding evolutionary and dynamical 
aspects from economic models a mathematical treatment often becomes possible; 
this however comes at the price that many fundamental features of economics will as a consequence not be understood, 
see e.g. \cite{farmer08}. 
Most complex systems, and  many aspects of economics in particular,  can not be reduced  to a few parameters -
without throwing the prince out of  {\em Hamlet}. 
Schumpeter himself criticized e.g. J.M. Keynes for proposing abstract models on the basis of insubstantial evidence and  
for freezing all but a few variables. Of the remaining ones, one could then argue that one caused another in 
a simple fashion. For this the expression {\em Ricardian vice} was coined \cite{schumpeter1954}, for an overview see \cite{richardian}. 
A dangerous consequence of this vice is that it suggests that valid policy conclusions could be derived. 
Note that what is today called Schumpeterian growth theory originated in the 
1990s \cite{romer1990,segerstrom1990,grossman1991,aghion1992,romer1994}. 
It relates output to labour, technology and population growth under various assumptions, and 
has little to do with the evolutionary aspects of  Schumpeterian economics.
Our following proposal of a dynamical, fully endogenous evolutionary model has nothing to do with these developments.  

Schumpeter's contributions on economic development are phrased in non-mathematical terms --
 for a good reason:  His ideas are non-equilibrium  concepts based non-linear models of evolutionary
processes, which are hard to (to some extent maybe impossible) capture in mathematical terms. 
Only in recent years there have been serious attempts to make evolution dynamics -- including their endogenous destructive 
elements (see e.g. \cite{newman}) -- a predictive and falsifiable theory, in the sense 
that quantitative models generate testable predictions on e.g. statistical features of evolutionary time series, such as the 
species diversity, species lifetimes, genera per species, etc. \cite{kauffman_NK,baksneppen,solemarubia,newman,raup,klimek08,thurner09}.
A large number of recent models are based on Kauffman networks, see e.g. \cite{kardanoff} and references therein. Statistical features, often characterized by power-laws in corresponding 
distribution functions, can then be compared with e.g.  fossil data. 
A particular experimental feature of evolutionary timeseries is the occurrence of so-called punctuated equilibria, which imply 
that the diversity of species is relative robust over large timescales,  and  only changes (often drastically) over 
very short timescales. This leads to  non-Brownian processes showing clustered volatility. 
For pioneering work on punctuated equilibria see \cite{farmer}.

\begin{figure}
\begin{center}
\includegraphics[width=4.8cm]{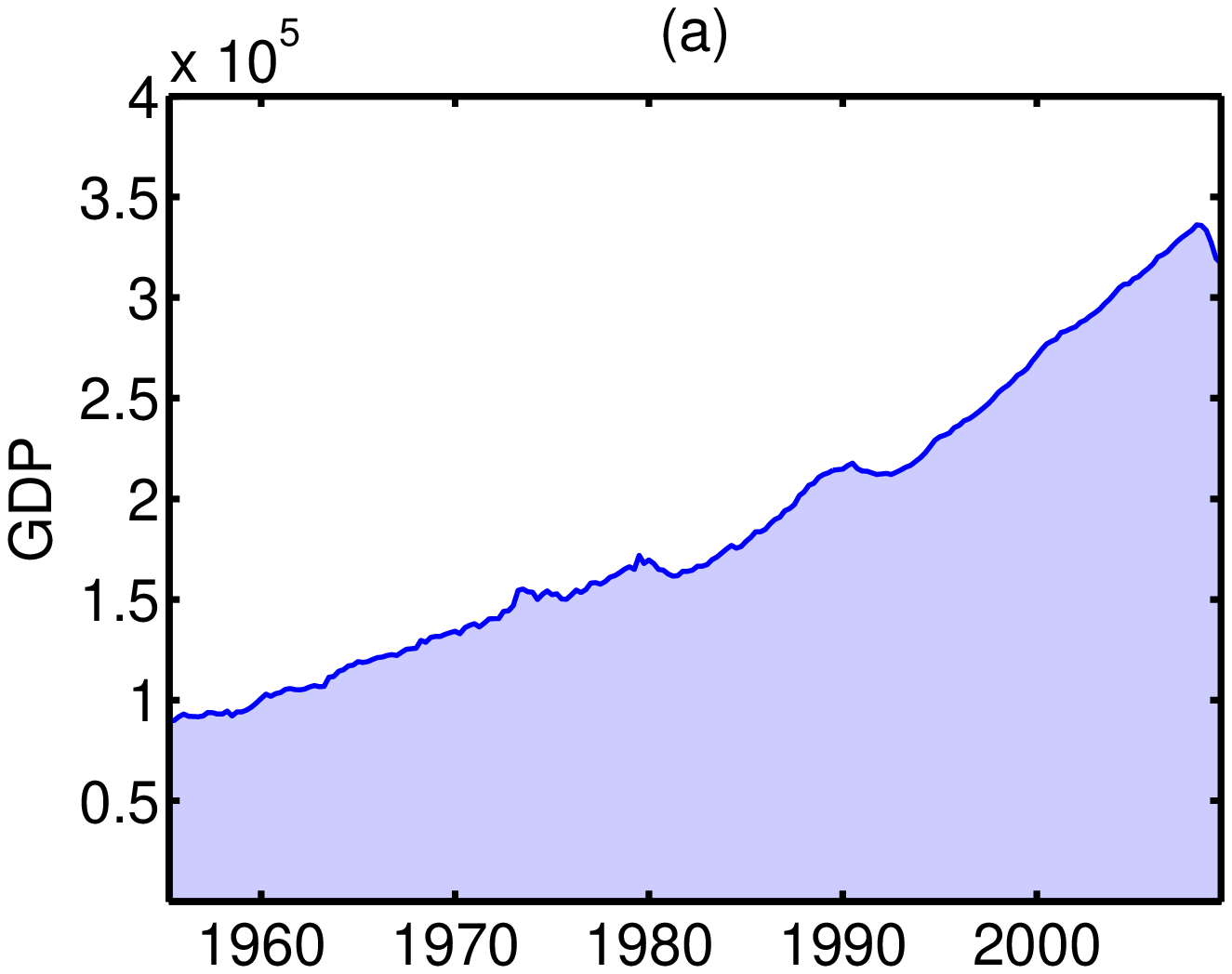}
\includegraphics[width=4.8cm]{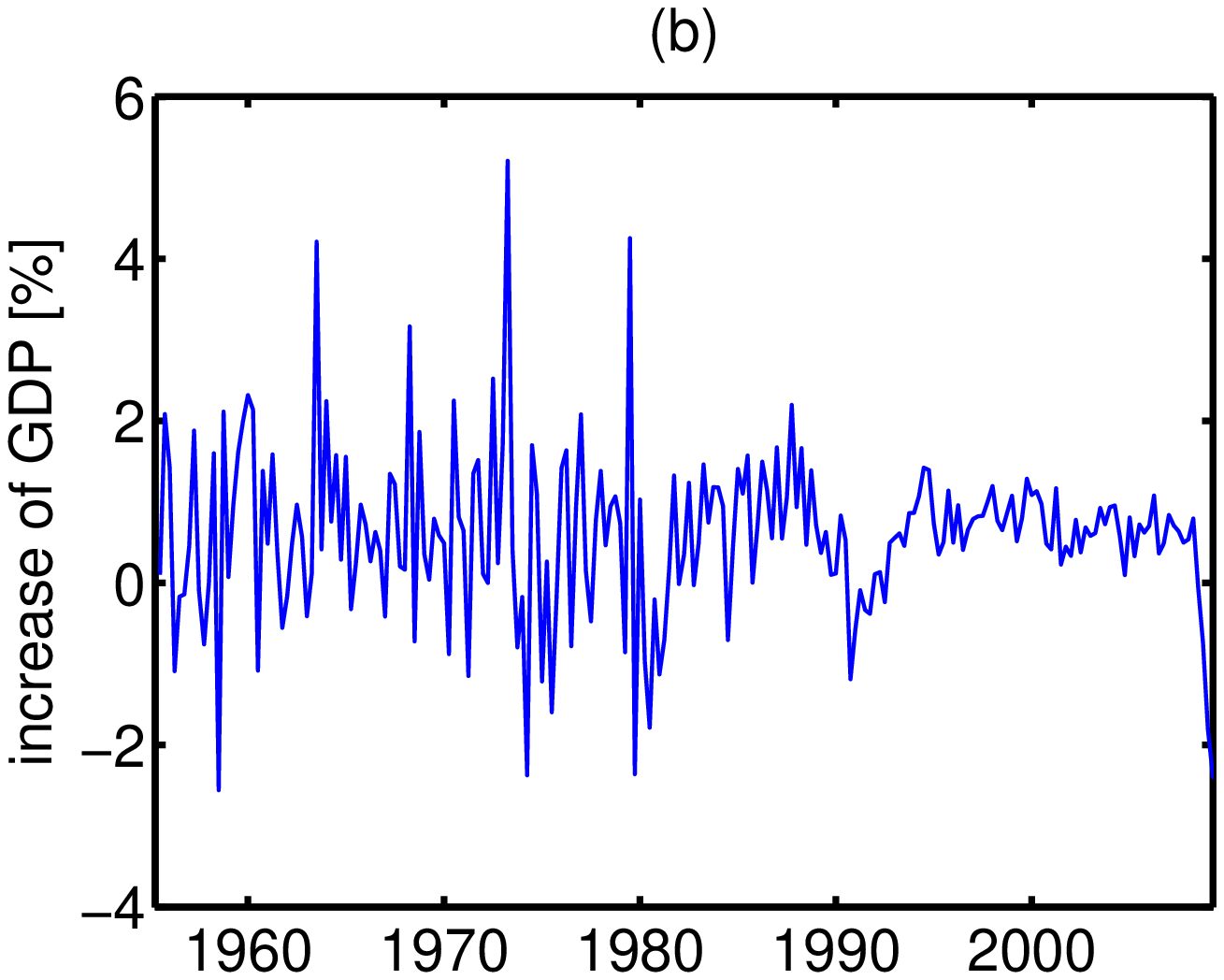}
\includegraphics[width=4.8cm]{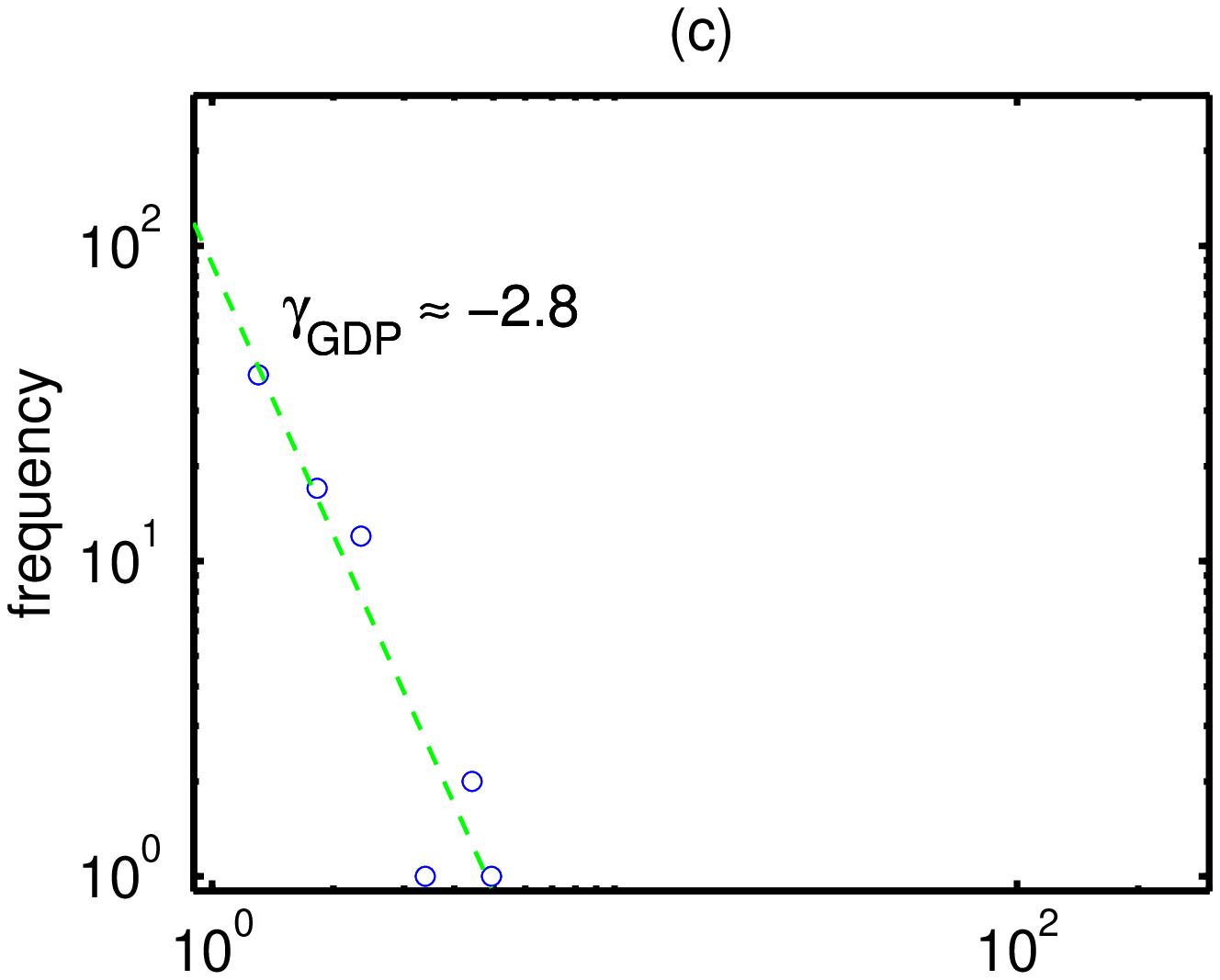} \\
\includegraphics[width=4.8cm]{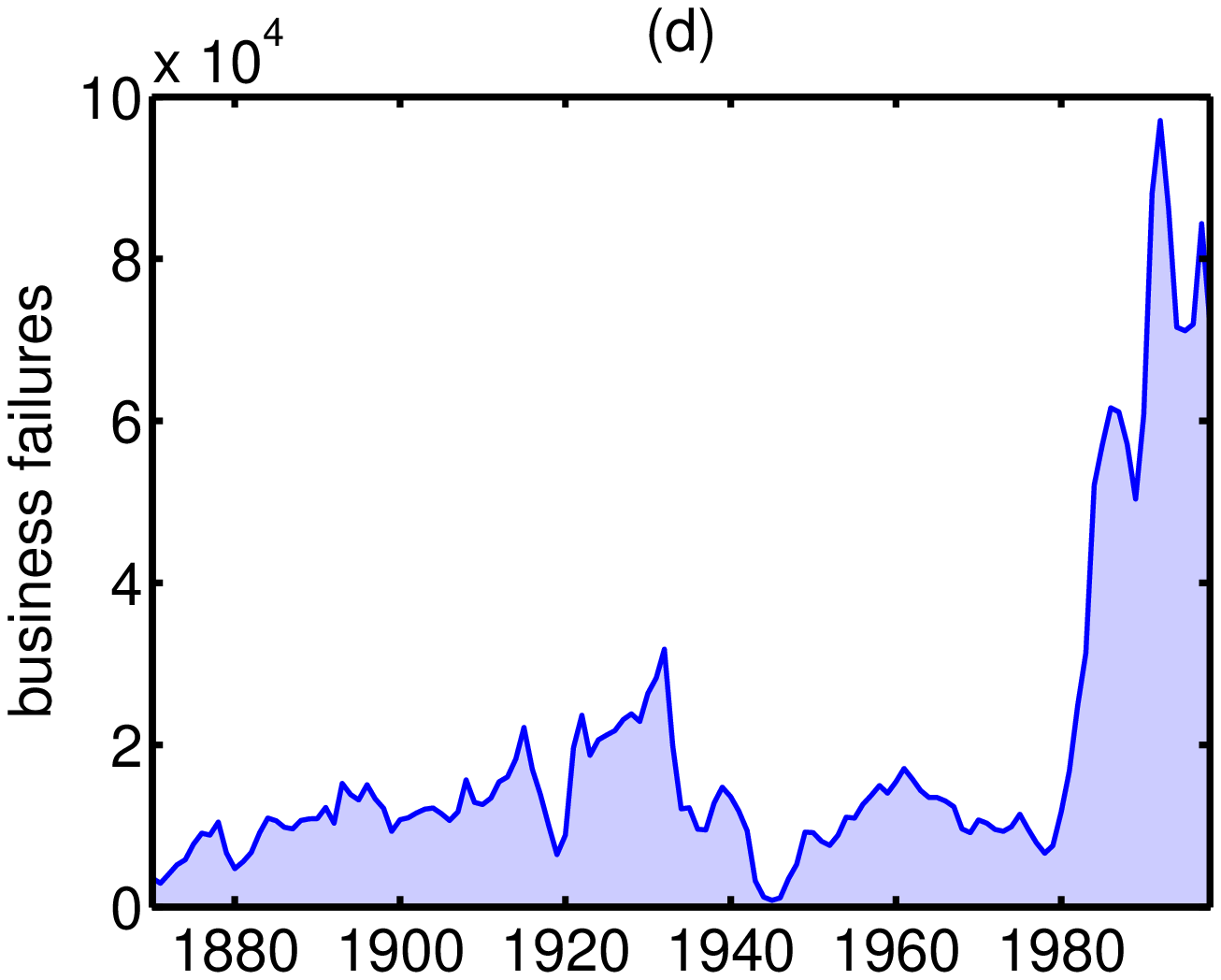}
\includegraphics[width=4.8cm]{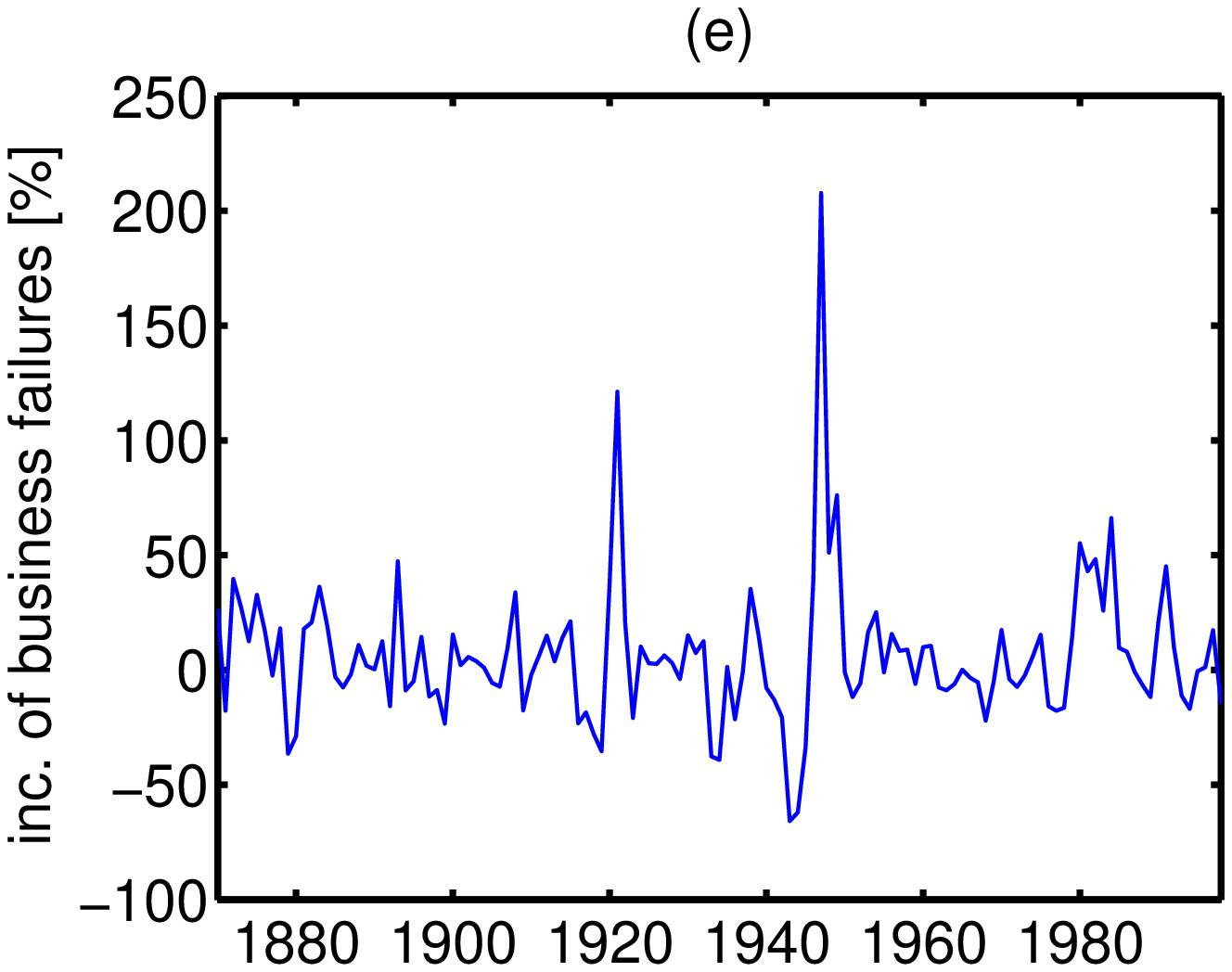}
\includegraphics[width=4.8cm]{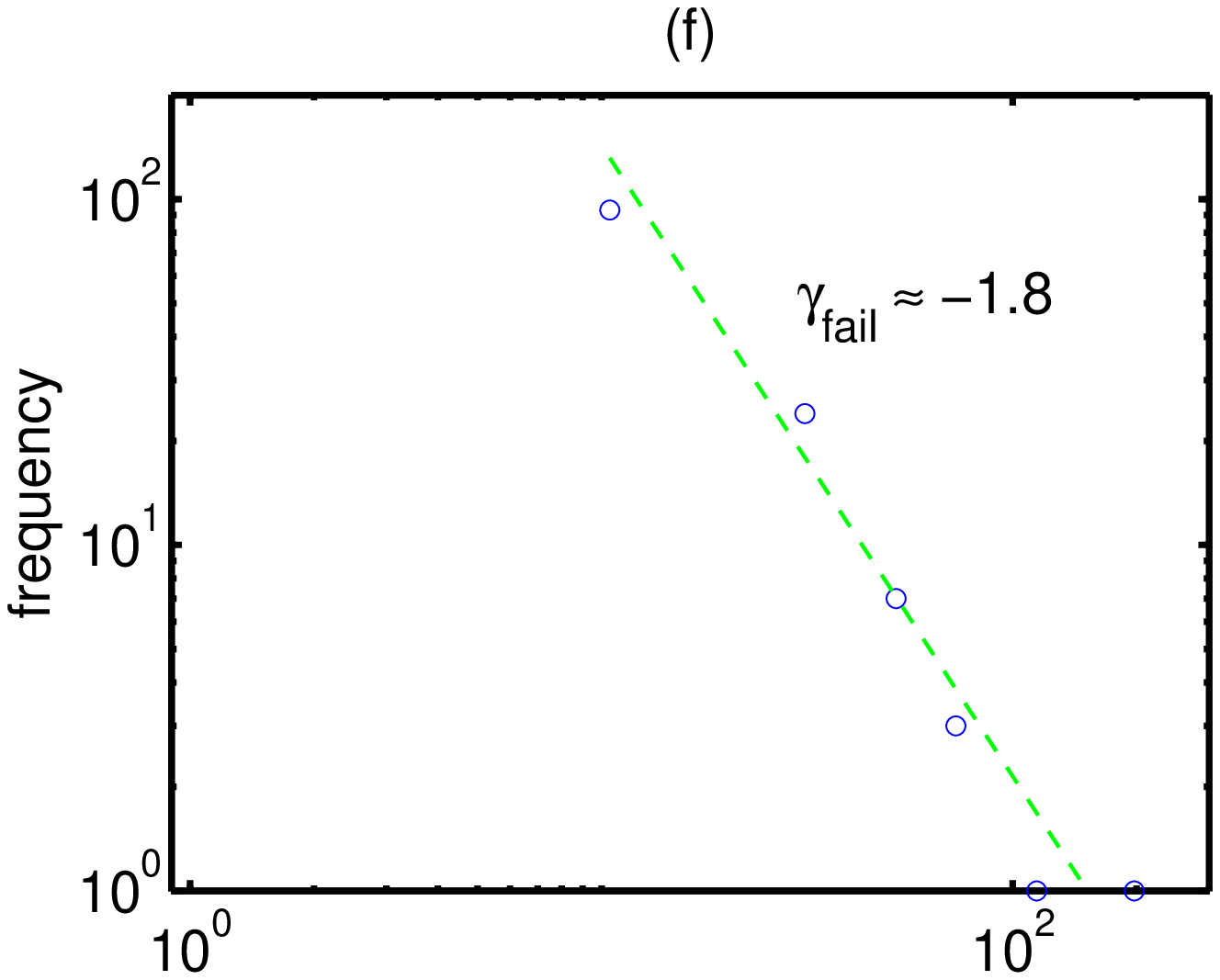} \\
\includegraphics[width=4.8cm]{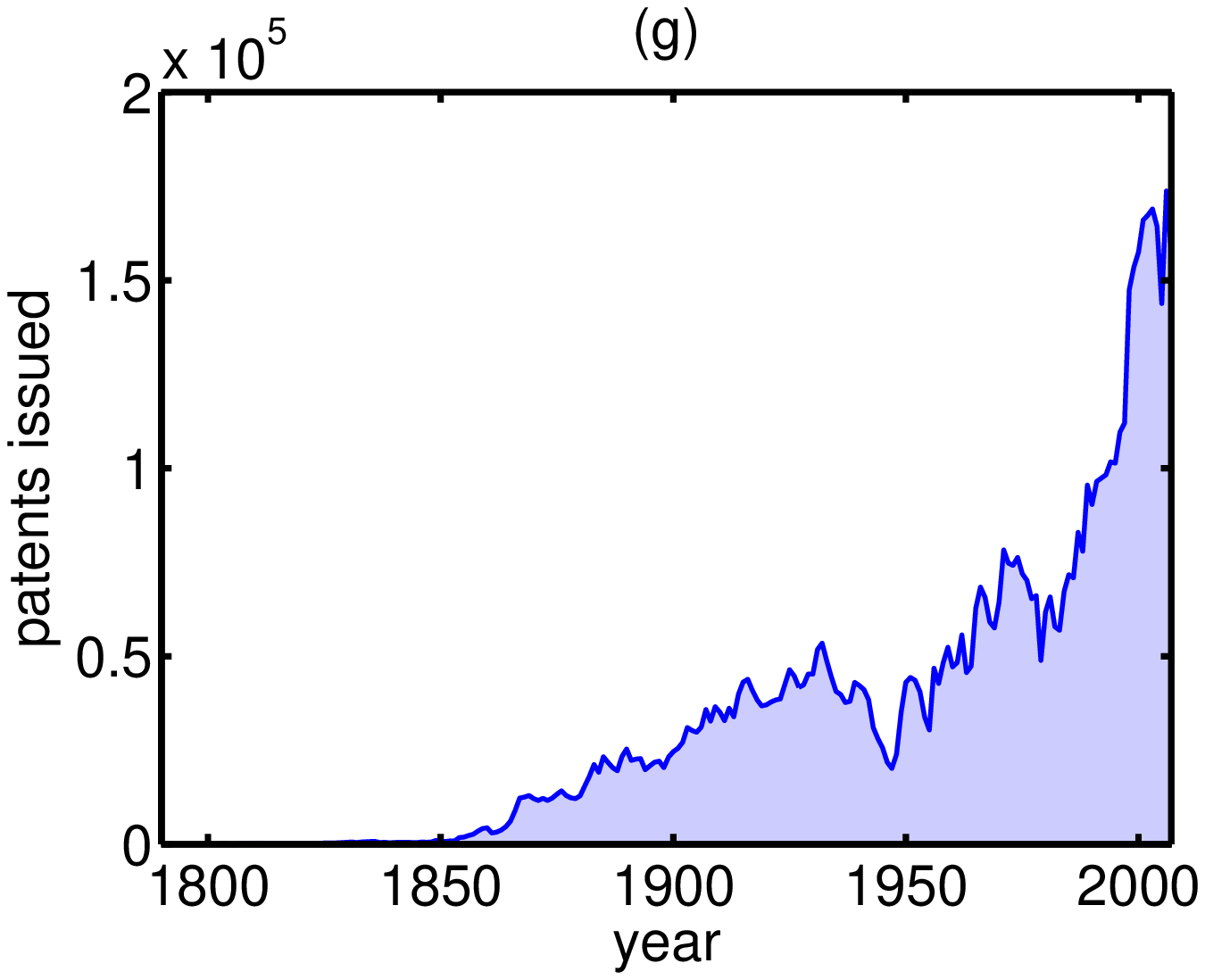}
\includegraphics[width=4.8cm]{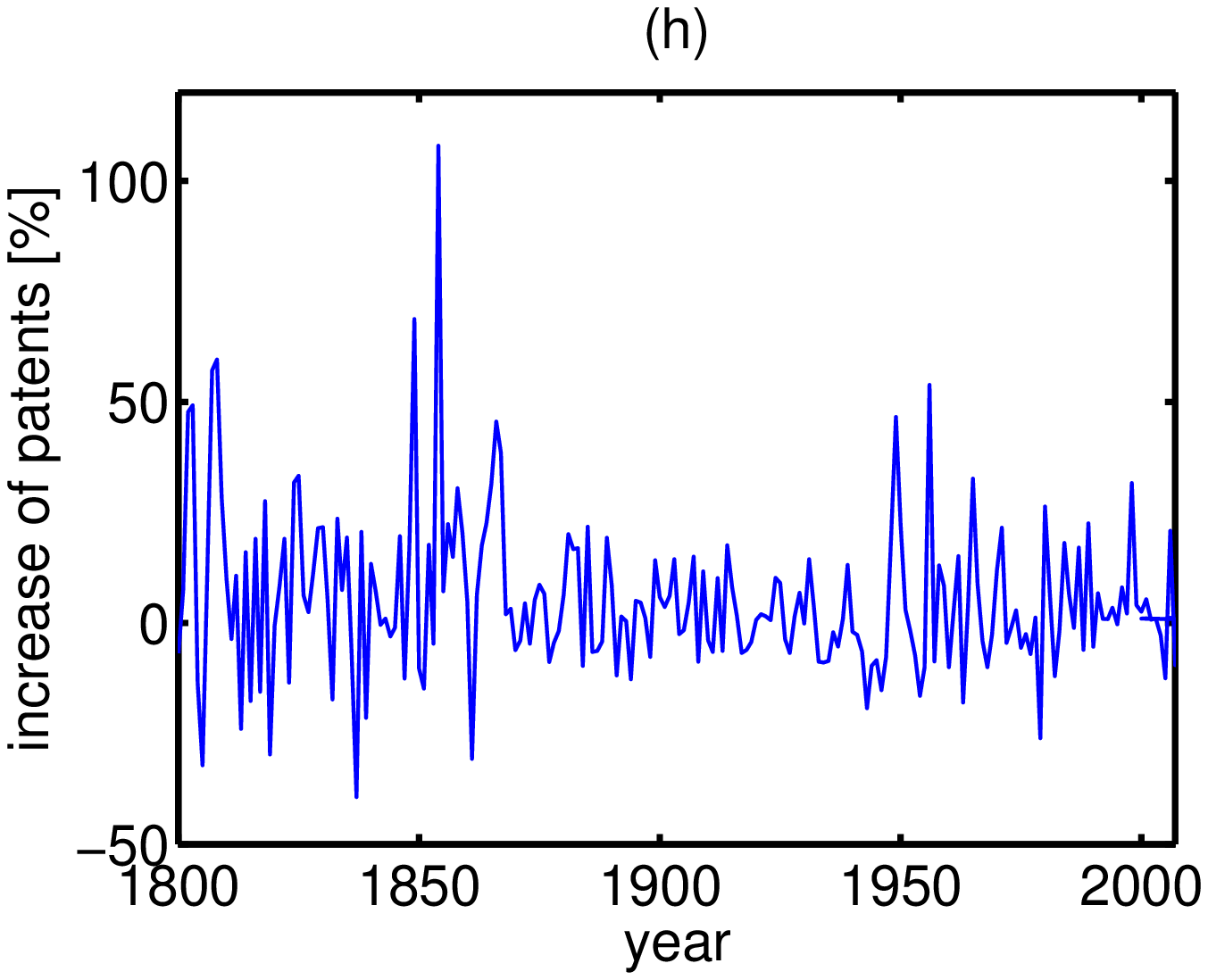}
\includegraphics[width=4.8cm]{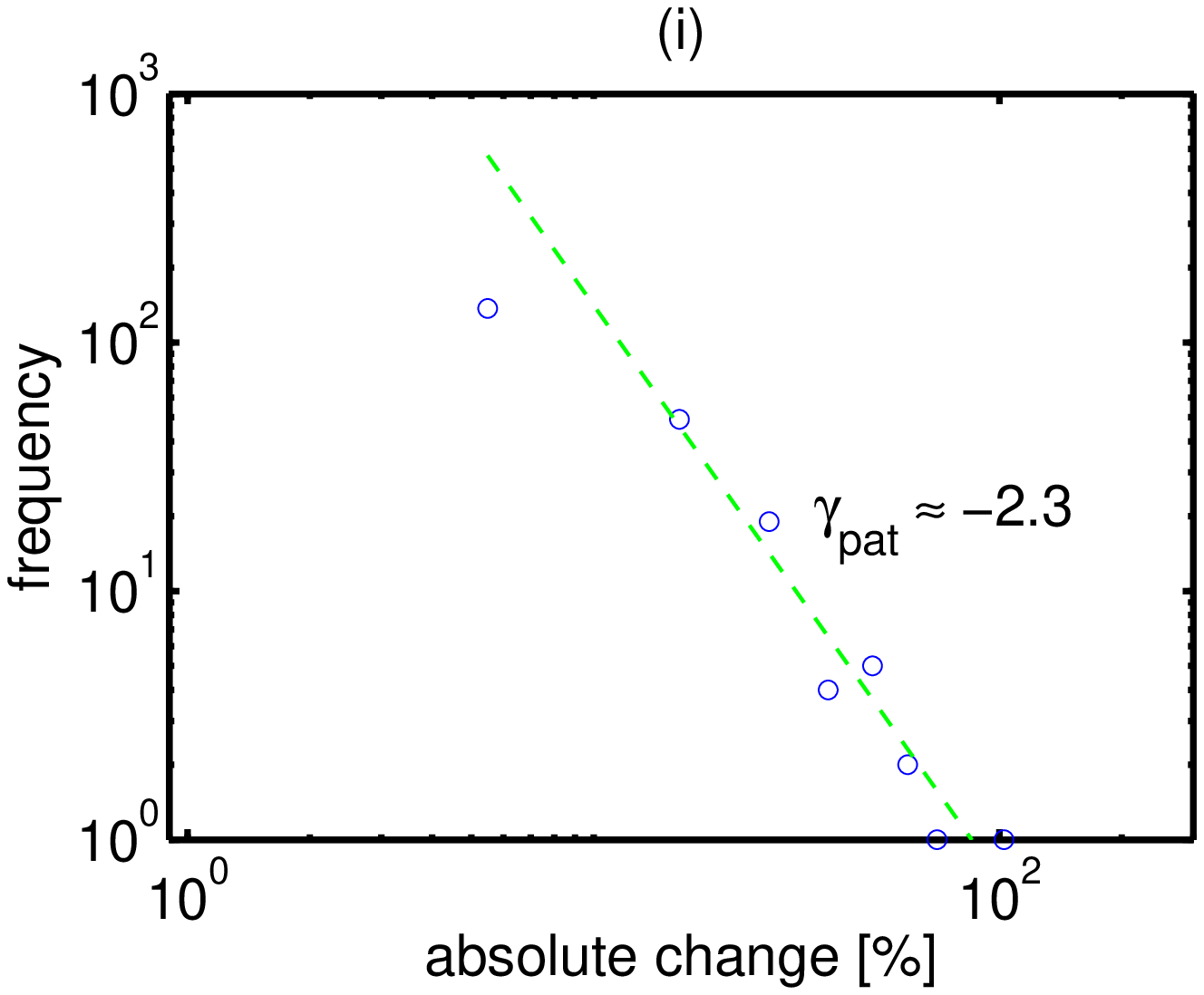} \\
\caption{
(a) GDP of the UK starting 1950 \cite{gdp}. 
(b) Percent increase of GDP from (a). 
(c) Histogram for (b); a least squares fit to a power-law yields a slope  of $\approx -2.8$.
%
(d)-(f) Number of business failures in the conterminous United States from 1870 onward, data from \cite{USCensus}. 
(e) Annual change in percent for (d), (f) histogram for (d);  power-law exponent $\approx -1.8$.
(g) Total number of patents issued on inventions in the United States from 1790 to 2007, data from \cite{USCensus}. 
(h) Annual increase of patents in percent, starting 1800. 
(i)  Histogram of absolute values of (h);  power-law  exponent $\approx -2.3$.
}
\end{center}
\label{timeseries}
\end{figure}

These features are also present in economic timeseries related to Schumpeterian dynamics. In Fig. \ref{timeseries}
we show timeseries for GDP (as a proxy for economic productivity), the number of firm failures (as an example of destructive 
dynamics) and the number of patents issued in the US   (for an estimate for the innovation potential). 
In all cases the percent increments of the timeseries show three characteristics: they show 
phases of relatively little activity followed by bursts of activity, they show clustered volatility, and 
they show non-Gaussian distributions, whose moments are summarized in Table \ref{tab1}. 

\begin{table}
\begin{center}
\caption{Moments of the percent-increment processes shown in Fig. \ref{timeseries} and model results for two choices of the 
model parameter $p$.}
\begin{tabular}{l c c c c}
                                   & mean & variance & skewness & kurtosis \\
\hline
GDP                             &  0.58  &     1.04   &  0.26  &  6.30 \\
business failures	&  6.23   &939.68   &  2.58  &17.49 \\
patents		         &  5.17  & 305.12   &  1.58  &  9.13 \\
model (p=0.01)		&  0.70  & 138.90   &  0.43  &   3.77 \\
model (p=0.0001)	&  0.21  &    40.11  &  0.81  &    9.63	 \\	
\end{tabular}
\end{center}
\label{tab1}
\end{table}

The bottleneck of a formal understanding of Schumpeterian processes is the current understanding of 
evolutionary dynamics. The reason why progress is limited in this direction lies in the mathematical and 
conceptual difficulties in dealing with this type of dynamics. In particular the facts that the system is an open system
(hard to find a 'constant' of motion, diversity, interactions, new possibilities constantly change). That the 
system does not reach an interesting or meaningful equilibrium is a direct consequence. 

Some recent progress in a quantitative formulation of the  dynamics of technological progress was made in 
\cite{barthur}, where certain products (logical circuits) are produced at random. Each of these products has a certain 
characteristic function (they compute something). According to their function these products get selected 
through an exogenous selection mechanism. Even though the model produces some realistic results in the spirit of Schumpeterian 
dynamics the necessity  of exogenous utility or fitness functions for the selection process is unsatisfactory. 
By trying to get rid of such endogenous elements in  describing evolutionary dynamics some purely endogenous 
models for creative and destructive processes have been proposed  \cite{htk1,htk2}. These however miss a  satisfactory 
combination of productive and destructive elements, which is attempted here. 

In this paper we present a simple toy model that tries to capture the essence of Schumpeterian dynamics. 
It is (in principle) an open model, that endogenously produces new goods and services. 
These new elements in the system then `interact' with each other  (and with old existing goods) in the sense 
that these interactions  can serve to aid the production of yet new goods and services. 
Which good can produce an other one is predetermined in an infinite hypothetical  {\em production table} , 
which captures all possible (thinkable) combinations of goods that lead to the production of novel ones. 
Also the opposite is true, a good that gets produced can have an adverse effect on an existing 
good. Which good -- given it gets produced -- drives out which other  good, is given by a 
predetermined {\em destruction table}.

This model builds on previous work developed in a biological context \cite{thurner09}.   
The aim of the present model is to provide a tool that allows to understand Schumpeterian 
dynamics, including its gales of destruction  within a minimum framework.
The model is able to provide an open, non-equilibrium concept and explains important dynamical facts of Schumpeterian 
dynamics. These include phases of boosts in economic development (measured in the diversity of available goods), 
phases of crashes, and phases of relative stability followed by turbulent restructuring of the 
entire economic 'world'. It is possible to interpret the successions of characteristic  
phases of construction, destruction and relative stability as 'business cycles'. 
We observe clustered volatility  and power-laws in our model data. Within this model we are able to understand various phases of
Schumpeterian dynamics as topological (emergent) properties of the dynamical production networks. 
Parts of the model are exactly solvable, for the full model however, we have to rely on a simple agent based computational realization.

\section{Model}

Schumpeterian economics is driven by the actions of `entrepreneurs' who realize  business ideas. 
The result of these actions are new goods and services which enter the 
economic scene --  the market. 
Usually new goods and services are (re)combinations or substitutions of existing things. 
The Ipod is a combination of some electrical parts, Wikipedia is a combination of the internet and the concept of an encyclopedia.  
New goods and services can `act on the world' in three ways: They can be used to produce other new things 
(as e.g. modular components), they can have a negative effect on existing things  by  suppressing their production or 
driving them out of the market  (destruction), or they have no effect at all.  
 
\subsection{Goods} 

In our simple model all thinkable goods and services are components in a time-dependent  $N$-dimensional vector $\vec \sigma(t)$. 
$N$ can be very large, even infinite. For simplicity the  components of this vector are binary. $\sigma_i(t)=1$ means that 
good $i$ is present (it exists) at time $t$, $\sigma_k(t)=0$, means service $k$ does not exist at $t$, either because 
it is not invented yet, or it got eliminated from  the market. (In a more general continuous setting, the state vector $0<\sigma_i(t)<1$,  
could be the relative abundance of good $i$ w.r.t. to the abundances of the other goods). 
New products come into being through combination of already existing products. 
An innovation -- the production of a new good $i$ -- can only happen if all necessary components (e.g. parts) are simultaneously available (exist). 
If a combination of goods $j$ and $k$ is a method to produce $i$, both $j$ and $k$ must be available in the system. 
Technically  $\sigma(t)$ can be seen as the  {\em availability status}, if $\sigma_i(t)=(0)1$ product $i$ is (not) accessible for production 
of future new goods, nor for the destruction of existing ones. 
The \emph{product diversity} of the market is defined as $D(t)= \frac1N \sum_{i=1}^N\sigma_i(t)$.

\subsection{Entrepreneurs/Production} 

\begin{figure}
\begin{center}
\includegraphics[height=7.8cm]{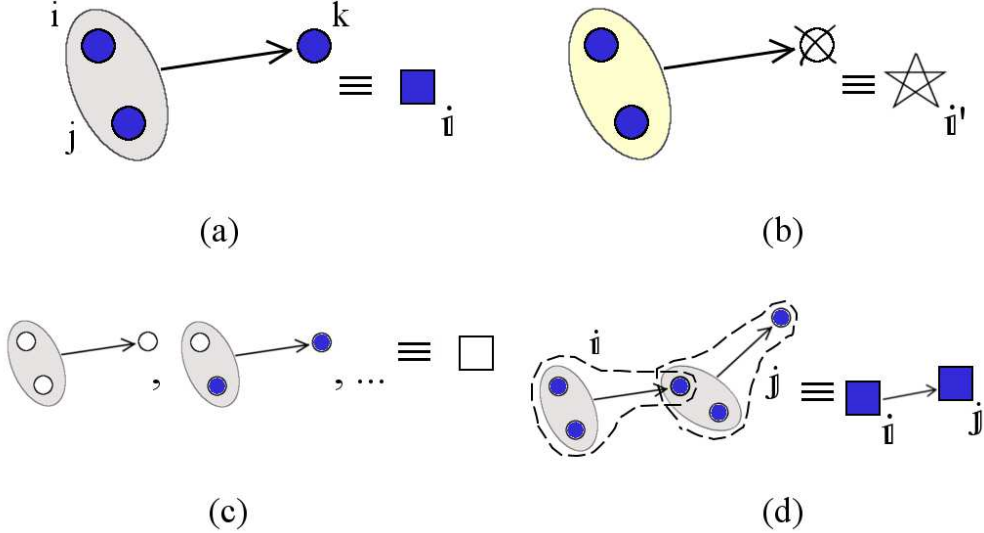}
\end{center}
\caption{
(a) Illustration of a production process. Products $i$ and $j$ reside in a productive set. 
There exists a production rule $\alpha^+_{ijk}=1$. Thus product $k$ becomes produced. This active production 
is depicted as a full square and indexed by {$\mathbb i$}. 
(b)  The same as (a) for a destruction process. Products $i'$ and $j'$ substitute $k'$ via the destruction rule $\alpha^-_{i'j'k'}=1$.
(c) Examples of  non-active productions (6 possible). Non-active productions are symbolized as open squares. 
(d) Definition of a link in the active production network: if a good produced by a production {$\mathbb i$} is part of the 
productive set of an other active production {$\mathbb j$}, production  {$\mathbb j$} gets a directed
link from production  {$\mathbb i$}. 
}
\label{prod_dest}
\end{figure}

Whether a product $k$ can be produced from components $i$ and $j$  is encoded in a \emph{production table}, $\alpha_{ijk}^+$. 
If it is {\em possible} to produce good $k$ from $i$ and $j$ (i.e. $\alpha^+_{ijk}>0$), this is called  a {\em production}. 
An entry in the  production table is in principle a real number which quantifies the {\em rate} at which a good is produced.
 Here for simplicity an entry in  $\alpha_{ijk}^+$ is assumed to be 
binary, 0 or 1. If goods $i$ and $j$ can produce $k$, $\alpha^+_{ijk}=1$,  goods $i$ and $j$ are called 
the {\em productive set} of $k$. If there is no production method associated with this combination, $\alpha^+_{ijk}=0$. 
The production process is then given by 
\begin{equation}
\sigma_k(t+1) = \alpha_{ijk}^+ \sigma_i(t) \sigma_j(t) \quad , 
\end{equation}
regardless whether $\sigma_k(t)=0$ or 1. If a production is {\em actually} producing $k$, 
(i.e. $\sigma_i(t)=\sigma_j(t)=\sigma_k(t) = \alpha^+_{ijk}=1$), we call it an {\em active production}, see Fig. \ref{prod_dest} (a).  

The role of the entrepreneur is to discover that $k$ can get produced as a combination of $i$ and $j$, 
i.e. to discover and activate the production. In general, a particular  good can be produced through more than one production method. 
In our (binary) notation the number of ways to produce good $k$ is $N_k^{\rm prod}(t)= \sum_{ij} \alpha_{ijk}^+ \sigma_i(t) \sigma_j(t)$.

\subsection{Competition/Destruction } 
 
 If a new product can get produced and
 serves a purpose (or a need) which hitherto has been  provided by 
 another product, the new and the old products are now in competition. The good that can be produced in a cheaper way 
 or that is more robust etc., will sooner or later drive the other one from the market. 
 This mechanism we incorporate in the model by allowing that a combination of two existing goods can lead 
 to a destructive influence on another product. The combination of goods $i'$  and $j'$ produces a good $l$ which 
 then drives product $k'$ out of the market. To keep it simple we say: the combination of $i'$ and $j'$ has a destructive 
 influence on $k'$.  We capture all possible destructive combinations in  a \emph{destruction rule table} $\alpha_{i'j'k'}^-$, 
 see Fig. \ref{prod_dest} (b). 
If $\alpha^-_{i'j'k'}=1$, products $i'$ and $j'$ substitute product $k'$.  We call $\{ i',j' \}$ the {\em destructive set} for  $k'$. 
Note that in this way we don't have to explicitly produce the competing good $l$. 
In the absence of a destruction process, $\alpha^-_{i'j'k'}=0$. 
As before an {\em active destruction} is only happening if $\sigma_{i'}(t)=\sigma_{j'}(t)=\sigma_{k'}(t) = \alpha^-_{i'j'k'}=1$. 
The  elementary dynamical update for a destructive process reads (for a good which is present at time $t$, i.e. $\sigma_k(t)=1$),
 \begin{equation}
 \sigma_k(t+1) = 1- \alpha_{ijk}^- \sigma_i(t) \sigma_j(t)
 \end{equation}
 In general , at any given time,  a good $k$ can be driven out of the market by more than one substitute -- in our notation -- by 
 $N_k^{\rm destr}(t)= \sum_{ij} \alpha_{ijk}^- \sigma_i(t) \sigma_j(t)$ destructive pairs.

Imagine the $N$ goods as circles assembled on a plane, see Fig. \ref{pro} (a). 
If they exist they are plotted as full blue circles, if they are not produced, they are white open circles. 
All existing goods have at least one productive set (pair of 2 existing (blue) circles); at time $t$ there exist $N_i^{\rm prod}(t)$ such sets. 
Many circles will in the same way assemble to form destructive sets, the exact number for node $i$ being  $N_i^{\rm destr}(t)$.
Now draw a circle around each productive and destructive set and connect each set with the good it is producing/destroying. 
The graph which is produced this way, Fig \ref{pro} (a) is the {\em economic web} \cite{sciam} of products. 
In this web in general every good  will be connected to several productive/destructive sets. 
To specify a dynamics we decide that if  there exist more production processes associated with  a particular good 
than there exist destructive processes associated with it, the good will be produced. 
If there are more destructive than productive sets associated with a good, it will not be produced, 
or it will get destroyed if it exists. If the number of productive and destructive sets for a good $i$ are the same, 
the state of $i$ will not be changed, i.e. $\sigma_i(t+1)=\sigma_i(t)$.  More quantitatively this reads
\begin{eqnarray}
N_i^{\rm prod}(t) > N_i^{\rm destr}(t) & \to & \sigma_i(t+1)=1 \nonumber \\
N_i^{\rm prod}(t) < N_i^{\rm destr}(t) & \to & \sigma_i(t+1)=0 \nonumber \\
N_i^{\rm prod}(t) = N_i^{\rm destr}(t) & \to & \sigma_i(t+1)=\sigma_i(t)  . 
\label{rule}
\end{eqnarray} 
Note that a production or destruction is only active if both goods in its production/destruction set are currently available. 
Thus changes in the status of a product possibly induce changes in the status of active production/destruction network. 

\subsection{The active production network}
\label{apn}

\begin{figure}[t]
\includegraphics[height=5.8cm]{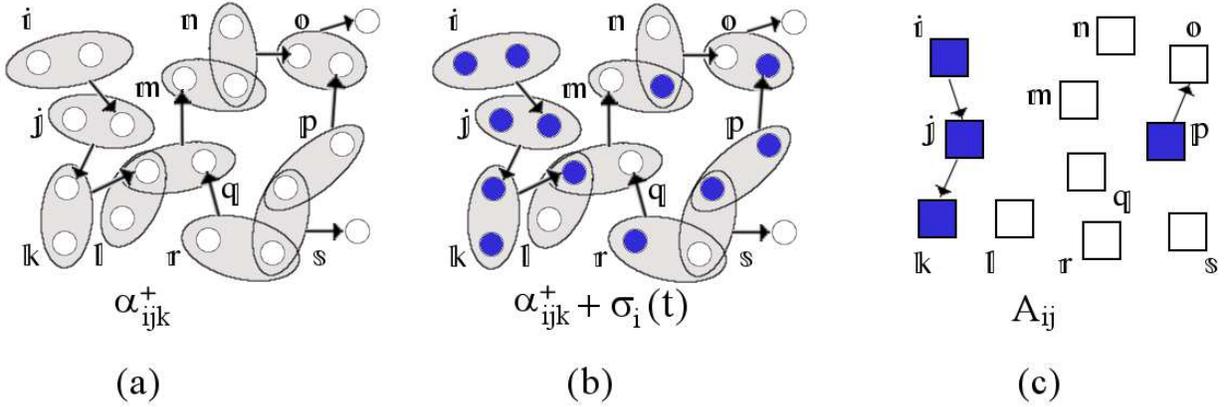}
\caption{
Comparison between static production rules $\alpha^+$ (a) and active production networks $A(t)$ (c). 
Production rules are all possible combinations to produce all thinkable goods in an economy. 
They correspond to all non-zero entries in $\alpha^{\pm}$. In (b) the actual state of the system $\vec \sigma(t)$
(which goods exist) is superimposed on $\alpha^{\pm}$. Representing (b) not in terms of goods but 
in terms  of productions we arrive at  the active production network (c), using the definition for links in Fig. \ref{prod_dest} (d).
}
\label{pro}
\end{figure}

It is essential to distinguish the production rules encoded in the tensors $\alpha^{\pm}$ and the 
{\em active production networks} $A(t)$. $\alpha^{\pm}$ is a collection of static rules of all potential ways to produce 
all thinkable goods. These rules exist regardless if goods exist or not. 
The production network $A(t)$ captures the set of actual active productions taking place at a given  time $t$.
It maps the state of the economy $\vec \sigma(t)$ (existing goods and services) with its rules $\alpha^{+}$
onto the set of active productions. 
It can be derived in the following way. 
A {\em production} is defined as a pair $(i,j)$ which produce a good $k$, and is nothing but a non-zero entry in $\alpha^{+}$.  
There are $N r^+$ productions in the economy, where $r^+$ is the (average) number of productions per good.
Non-existing goods are open circles, the symbol used for a production is a square.
A production is called an   {\em active production} if the  production set and the produced node all exist 
($\sigma_i(t)=\sigma_j(t)=\sigma_k(t)=1$).
An active production is shown in  Fig. \ref{prod_dest} (a) symbolized as a filled square. 
In (c) we show some examples of non-active productions (open square). 
We label active productions by bold-face indices, ${\mathbb i} \in \{1,\dots N r^+\}$.
These constitute the nodes of the {\em active production network}.
A directed link from active production node ${\mathbb i}$ to node  ${\mathbb j}$ is defined if
the {\em node} produced by production  ${\mathbb i}$ is in the productive set of production  ${\mathbb j}$, see 
 \ref{prod_dest} (d). It is then denoted as 
$A_{{\mathbb i} {\mathbb j}}=1$. This definition is illustrated in Fig. \ref{prod_dest} (c).
As an example of how to construct the active production network $A(t)$ from $\vec \sigma(t)$ and $\alpha^{+}$, see Fig. 
\ref{pro}. 
In Fig. \ref{pro} (a) we show a section of the static $\alpha^+$, in (b) we superimpose the knowledge of which of these nodes actually exist 
at time $t$. In (c) all productions are shown as squares (active ones full, non-active ones empty). The links between the active 
productions constitute the active production network. 
In this way we map the production rule tensor $\alpha^+$ onto a production (adjacency) matrix $A(t)$. It is  
defined on the space of all productions  and links two nodes 
if one node is the product of an active production currently fed by another node.
The active destruction network is obtained in the same way. 

After having constructed the active production network, we then remove all unconnected nodes. 
To detect {\em dominant}  links in this network, we  introduce the following threshold: we  
remove all links from  the active production network which exist less than a prespecified percentage of times 
$h$ within  a moving time-window of length $T$.  So if $h=95$ and $T=100$, the network at time $t$, $A(t)$, only contains links which have 
existed more than  95 times within the timewindow $[t-T,t]$.

\subsection{Spontaneous ideas and disasters}

From time to time spontaneous ideas or inventions take place, without the need of 
the production network.  Also from time to time goods disappear from the economy, 
say through some exogenous events. To model these we introduce a probability $p$
with which a given existing good is spontaneously annihilated, or a non-existing good 
gets spontaneously invented.   
This is the only stochastic component of the model (besides the update sequence) and 
has an important effect as a driving force. 

\subsection{A Schumpeterian algorithm}

Consider you are at timestep $t$, the update to $t+1$ happens in the following  three steps:
\begin{itemize}
\item pick a good $i$ at random (random sequential update)
\item sum all productive and destructive influences  on $i$, i.e. compute 
$\Delta^{\pm}_i (t) \equiv \sum_{j,k=1}^N (\alpha^+_{ijk}-\alpha^-_{ijk}) \sigma_j(t) \sigma_k(t).$ 
 If $\Delta^{\pm}_i (t)>(<)0$ set $\sigma_i(t+1)=1(0)$. 
  For $\Delta^{\pm}_i (t)=0$ do not change, $\sigma_i(t+1)=\sigma_i(t)$
 \item  with probability $p$ switch the state of $\sigma_i(t+1)$, i.e.  if $\sigma_i(t+1)=1(0)$ set it to  $\sigma_i(t+1)=0(1)$
 \item continue until all goods have been updated once, then go to next timestep 
\end{itemize}
As initial conditions ($t=0$) we chose  a fraction of randomly chosen initial goods, typically we set $D(0)\sim0.05-0.2$. 

In principle it is possible to empirically assess production or destruction networks in the real economy, 
however in practice this is unrealistic and would involve 
tremendous efforts. For a systemic understanding of Schumpeterian dynamics a detailed knowledge of these networks 
is maybe not necessary, and a number of statistical descriptions of these networks would suffice. 
The simplest implementation of a production/destruction network  is to 
use  random networks, i.e. to model $\alpha^{\pm}$ as a random tensors. These tensors can then be described by a single number 
$r^{+}$ and $r^{-}$ which are the constructive/destructive rule densities. With other words the probability that 
any given entry in $\alpha^+$ equals 1 is  
$P(\alpha_{ijk}^+=1) = r^+ {N \choose 2}^{-1}$, or 
each product has on average $r^{\pm}$ incoming productive/destructive links from productive/destructive sets.
Further, which goods form which productive sets is also randomly assigned, 
i.e. the probability that a given product belongs to a given productive/destructive set is $2 r^{\pm}/N$ (for $r^{\pm} \ll  N$). 

Certainly real production networks carry structure and logic;  the assumption that production networks 
are unstructured is unrealistic to some degree. 
For this reason we will look at scale-free versions of production/destruction topologies. 
Finally,  note that $\alpha^{\pm}$ is fixed throughout the simulation.

\begin{table}
\begin{center}
\caption{Summary  of model parameters.}
\begin{tabular}{lll}
Variable &  \\
\hline
$\sigma_i(t)$ & state of good $i$. exists / does not exist   & dynamic\\
$D(t)$ & diversity at time $t$ & dynamic\\
$A(t)$ & active production network & dynamic\\
& & \\
Parameter &  \\
\hline
$\alpha^{\pm}$ & productive/destructive interaction topology & fixed\\
$r^{\pm}$ & rule densities & fixed\\
$p$ & spontaneous-innovation parameter & fixed\\
\end{tabular}
\end{center}
\label{table2}
\end{table}

\section{Results}

\begin{figure}[t]
\begin{center}
\includegraphics[width=14.0cm]{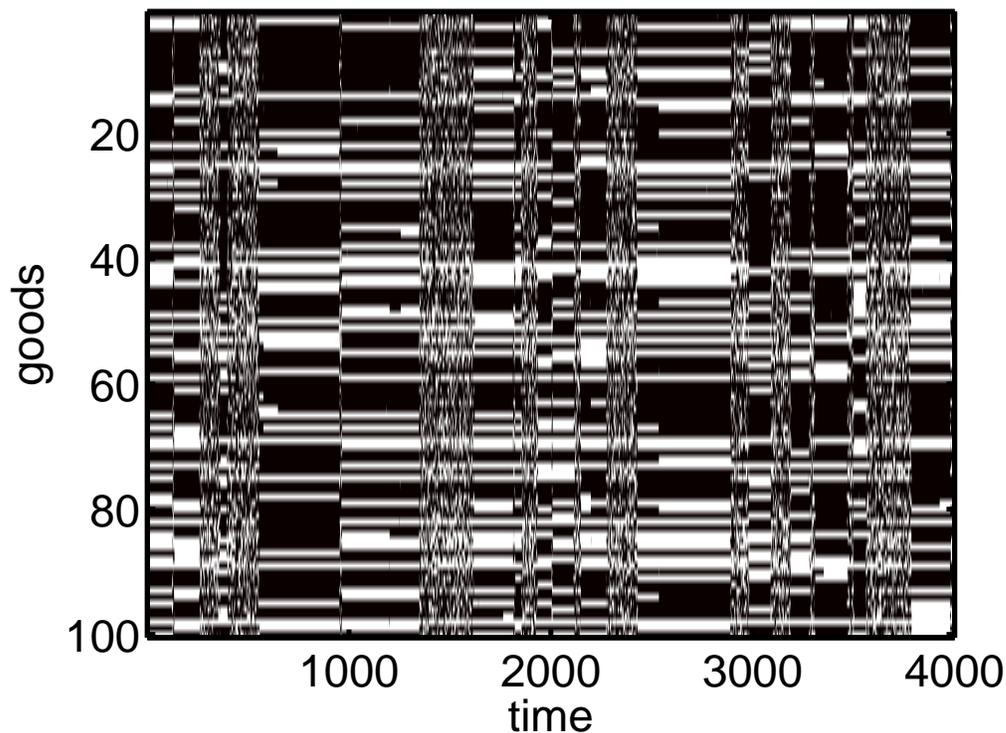}
\end{center}
\caption{Individual trajectories of all goods for the 
parameter settings $r^+=10$, $r^-=15$, $p=2 \cdot 10^{-4}$, $N=10^2$ and an initial diversity of $20$ 
randomly chosen goods. 
 }
\label{trajectory1}
\end{figure}

\begin{figure}[t]
\begin{center}
\includegraphics[width=7.3cm]{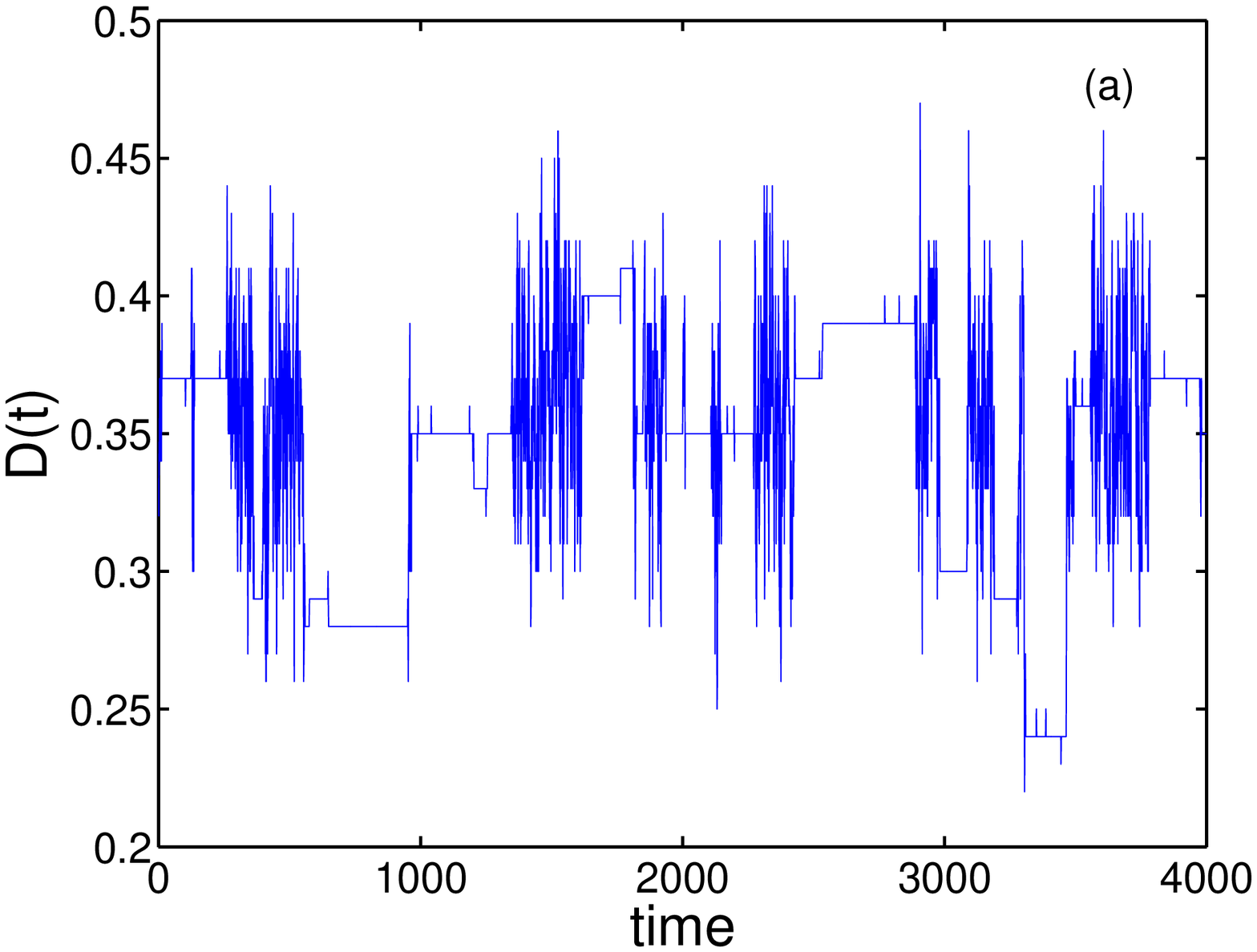}
\includegraphics[width=7.0cm]{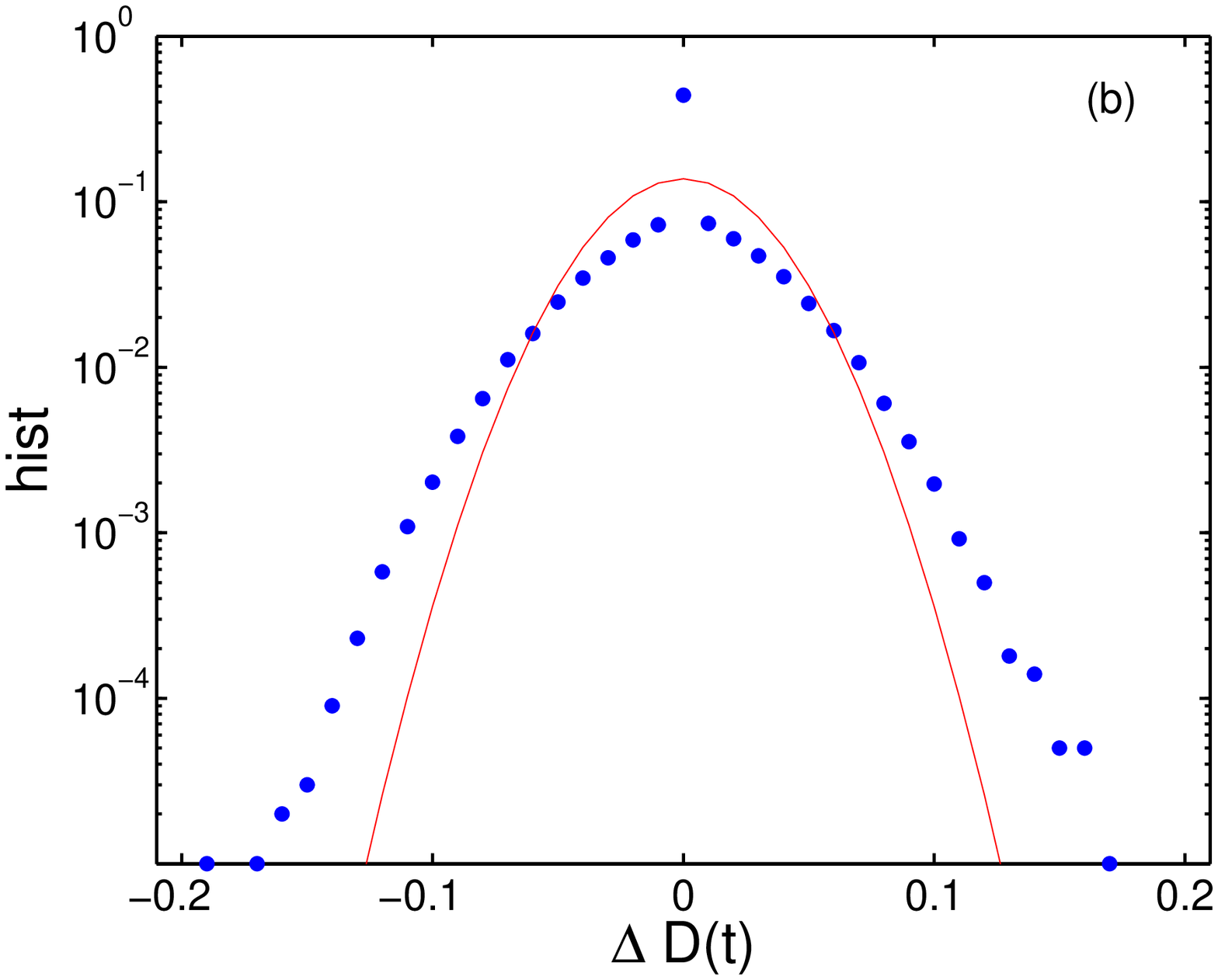}
\end{center}
\caption{
(a)  Diversity $D(t)$ from the simulation in the previous figure which shows a punctuated pattern. 
The system jumps between phases of relatively few changes -- the plateaus -- and chaotic restructuring  phases. 
The length of the chaotic phases is distributed as a power law which is identical with the fluctuation lifetime distributions 
shown in Fig. \ref{ClustSizes}. 
(b) Histogram over the percent-increments of diversity. The line is a Gaussian distribution with the same mean and variance.}
\label{trajectory}
\end{figure}

Implementing the system in a computer model we get a model dynamics as seen in Fig. \ref{trajectory1}, 
where we show the individual trajectories of 100 nodes. 
Time progresses from left to right. Each column shows the state of each of the goods $i=1, \dots , N$ at any given time. 
If good $i$ exists at $t$, $\sigma_i(t)=1$ is represented as a white cell,   a black cell at position $(i,t)$ indicates $\sigma_i(t)=0$. 
It is immediately visible that there exist two distinct modes in the system's dynamics, a quasi stationary phase, where the set 
of existing  products does practically  not change over time, and a phase of massive restructuring. 
To extract the timeseries of diversity $D(t)$ we sum the number of all white cells within one column at time $t$ and divide by 
$N$ which is shown in Fig \ref{trajectory} (a). Again it is seen that the plateaus of  constant product diversity (punctuated equilibria)
are separated by restructuring periods, characterized by large fluctuations of the products that exist. 
Note that quasi stationary plateaus differ in value. Depending on parameter settings, stationary diversity levels may differ 
by up to 50\%. 
These fluctuations are by no means Gaussian, as can be inferred from Fig. \ref{trajectory} (b), where the histogram of the 
percent-changes in the diversity timeseries, $R_t=(D(t)-D(t-1))/D(t-1)$, is shown. 
The skew in the distribution is absent when looking at the increment distribution of $D(t)$. 
For this parameter setting the dynamics of the system does not reach a static or frozen state, stationary phases and chaotic ones continue to 
follow each other.  We have checked this up to  $10^6$ simulation timesteps. 

For the number of active productions (a proxy for our model `GDP') and destructions (model `business failures') at every timestep we fitted the 
corresponding (percent increment) histograms to power-laws (not shown). The exponents are $\approx -2.6$ for the productions and $\approx -2.8$
for the destructions, respectively. While the exponent for productions coincides well with the one estimated from the GDP (Fig. \ref{timeseries} (c)), the destructive 
one is larger than for the business failures, Fig. \ref{timeseries} (f). 


\begin{figure}[t!]
\begin{center}
\includegraphics[width=16.0cm]{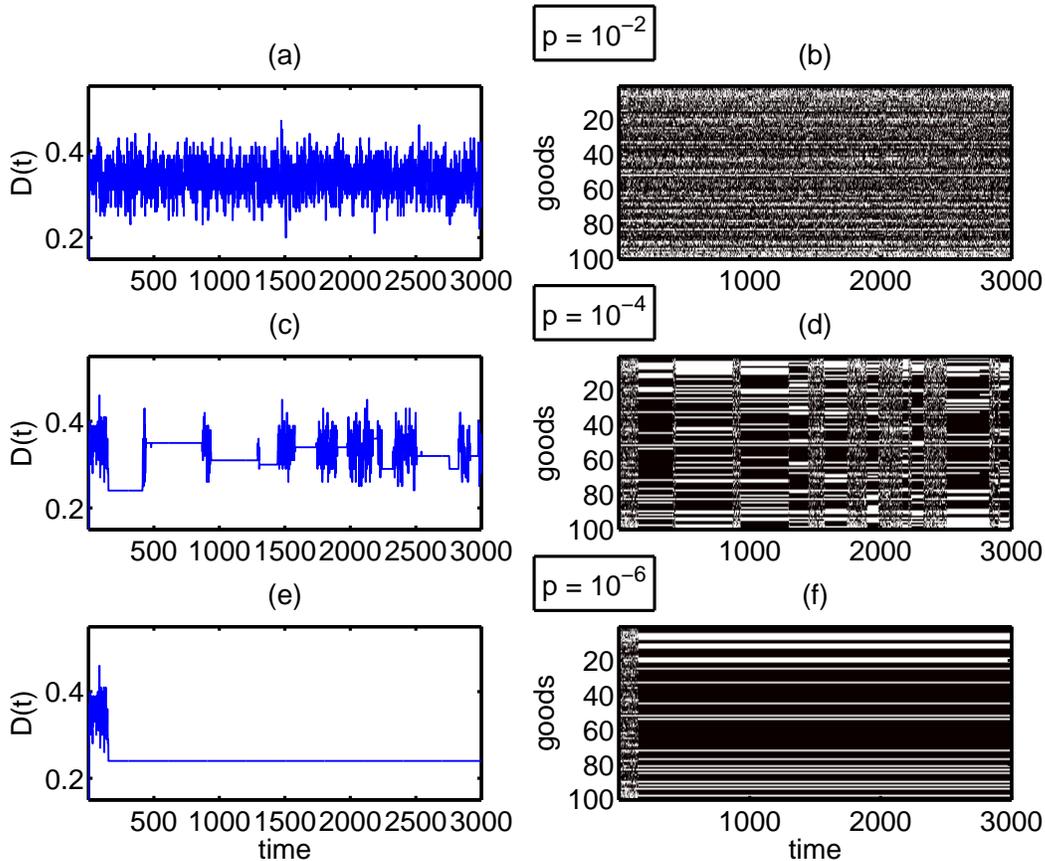}
\end{center}
\caption{Timeseries of goods and system diversity as in the previous figures for various values of $p$. 
For high innovation rates the system never settles into plateaus, $p=0.01$ (a)-(b), for intermediate levels $p=10^{-4}$ plateaus form (c)-(d), 
and for low levels $p=10^{-6}$ the system freezes, (e)-(f).}  
\label{drive}
\end{figure}

However, the dynamics changes with altering the `innovative' rate  $p$, see Fig. \ref{drive}. 
In the situation where there is a rate of $p=0.01$, (a)-(b), i.e. in a system of $N=100$ there is about one 
spontaneous innovation or destruction per timestep, we observe extended restructuring processes, almost never leading to plateaus.
For $p=10 ^{-4}$ (one innovation/destruction every 100 timesteps) the situation is as described above, (c)-(d). When $p$ gets too small 
($p \ll N^{-1}$) 
the system is eventually not driven from a stationary state and freezes (e)-(f). 
In the next section we will discuss alternatives to the driving process, where innovation rates $p$ get replaced by {\it product lifetimes}
or alternative competition models. These alternatives also drive the system dynamically away from frozen states. 

We have  studied under which topological circumstances the generic dynamics is maintained for $p$ in the range of $[10^{3}-10^{5}]$.
 If there are too many destructive influences w.r.t. constructive ones ($r^+ \ll  r^-$) the system will evolve toward a state of 
 low diversity in which innovations are mostly suppressed. If there are much more constructive than destructive interactions 
($r^- \ll  r^+$), i.e. little  competition, the system is expected to saturate in a highly  diverse state. 
In between these two extremal cases we find sustained  dynamics as described above. This regime is indeed very broad, 
and does not need a finetuning of $r^+$ and $r^-$.

\begin{figure}
\begin{center}
\includegraphics[height=6.4cm]{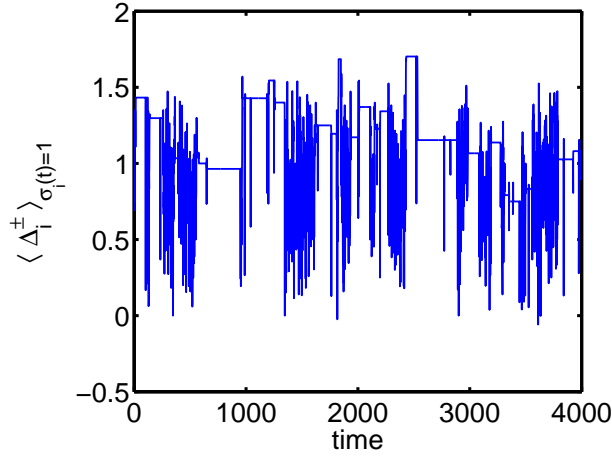}
\end{center}
\caption{Difference between active production and destruction influences per node $\langle \Delta^{\pm}_i \rangle_{\sigma_i(t)=1}$,  
averaged over all active nodes for the same run displayed in Fig. \ref{trajectory}. 
}
\label{delta}
\end{figure}

Let us finally give an intuitive explanation of what drives the dynamics of the system.  
The dynamics can be interpreted as if products tend to populate locations in product-space which are locally 
characterized by high densities of  productive rules and low densities of destructive influences. If the system remains 
in such a basin of attraction this results in diversity plateaus. 
Small perturbations can force  the activated population  of goods out of these basins. Products 
undergo a restructuring phase until an other basin of attraction  is found. 
 To quantify this Fig. \ref{delta} shows  
 $\langle \Delta^{\pm}_i \rangle_{\sigma_i(t)=1}$, averaged over all active nodes, for the
 simulation with time series shown in Fig. \ref{trajectory}).
This quantity indicates the surplus of production over destruction rules.  
Stable phases in diversity (plateaus in diversity in Fig. 
\ref{trajectory}) 
are always associated 
with higher values of $\langle \Delta^{\pm}_i \rangle_{\sigma_i(t)=1}$ than in the restructuring (chaotic) phases. 
Typical values in the stable phases lie between one and two, meaning that the average node can lose 
one productive pair and still be sustained. 
In the chaotic phases $\langle \Delta^{\pm}_i \rangle_{\sigma_i(t)=1}$ tends toward zero. 
Note here that once a node is deactivated it is not considered in the average anymore. This is the reason  
why $\langle \Delta^{\pm}_i \rangle_{\sigma_i(t)=1}\geq0$.

\section{Results on model variants}

\subsection{More realistic competition}
\label{alternat}

We studied the influence of hierarchical suppression as  a more realistic mechanism of competition. 
We start again with productive rules which are distributed randomly as above. Then for each node, we identified 
its productive set. 
If for example sets $\mathbb i, \mathbb j, \mathbb k$ all produce node $l$ we randomly assign a hierarchy 
on these sets, say $\mathbb j \rightarrow \mathbb i \rightarrow \mathbb k$, meaning $\mathbb j$ dominates $\mathbb i$, and $\mathbb i$ dominates $\mathbb k$. 
This domination could mean for example $\mathbb j$ produces $l$ cheaper than $\mathbb i$, and $\mathbb k$ is the most expensive way to assemble $l$. 
The intuition behind this approach is that if a more efficient way to produce node $l$ is available, as here for example  
the production $\mathbb i$  for good $l$ is cheaper than production $\mathbb k$, and $\mathbb i$ 
will supersede the old one ($\mathbb k$) and thus suppress its productive  set. 
If we find an even  better rule e.g. $\mathbb j$, this again suppresses the less efficient  productions $\mathbb i$ and $\mathbb k$. 

\begin{figure}[t!]
\begin{center}
\includegraphics[width=5.0cm]{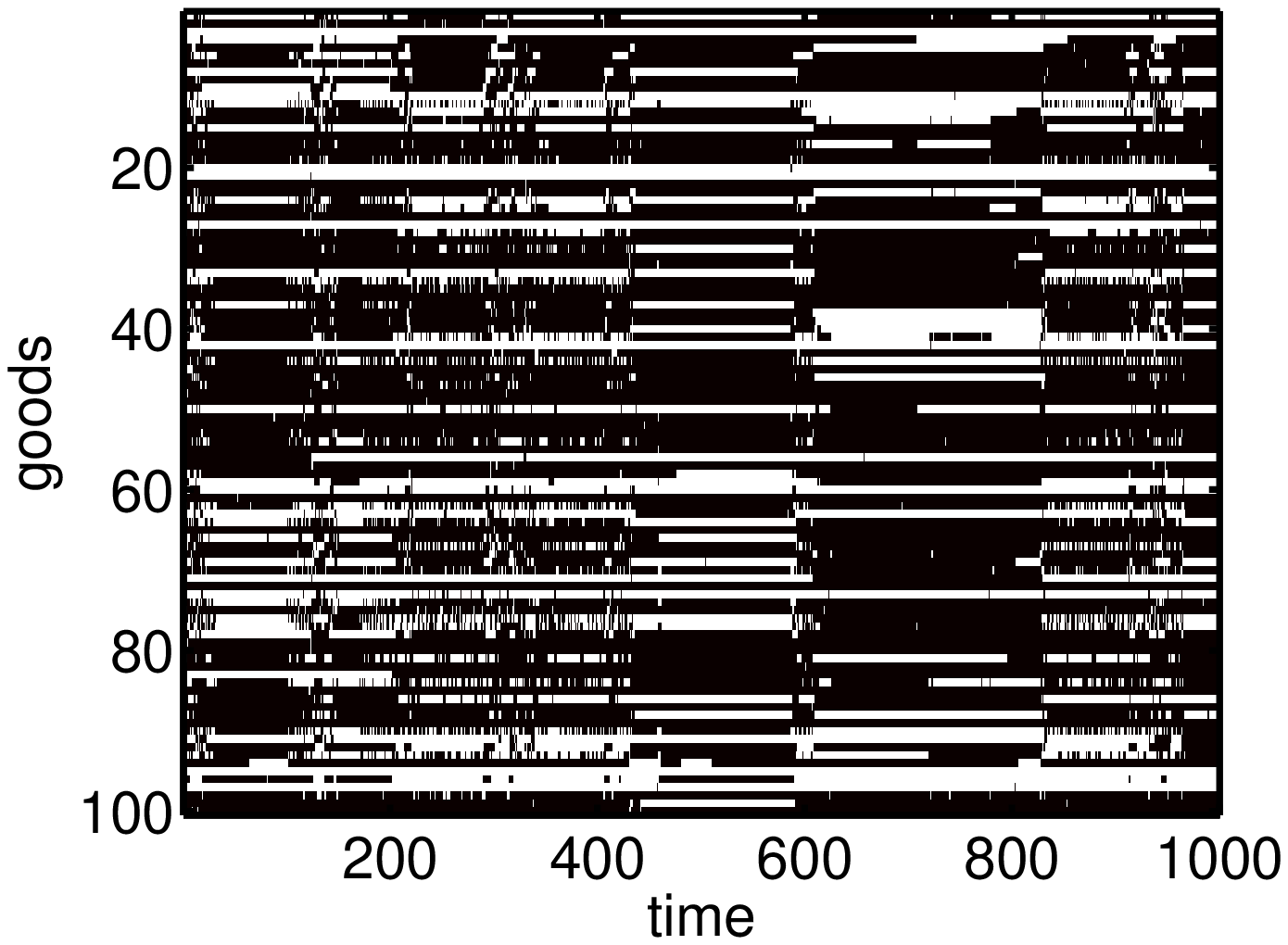}
\includegraphics[width=5.0cm]{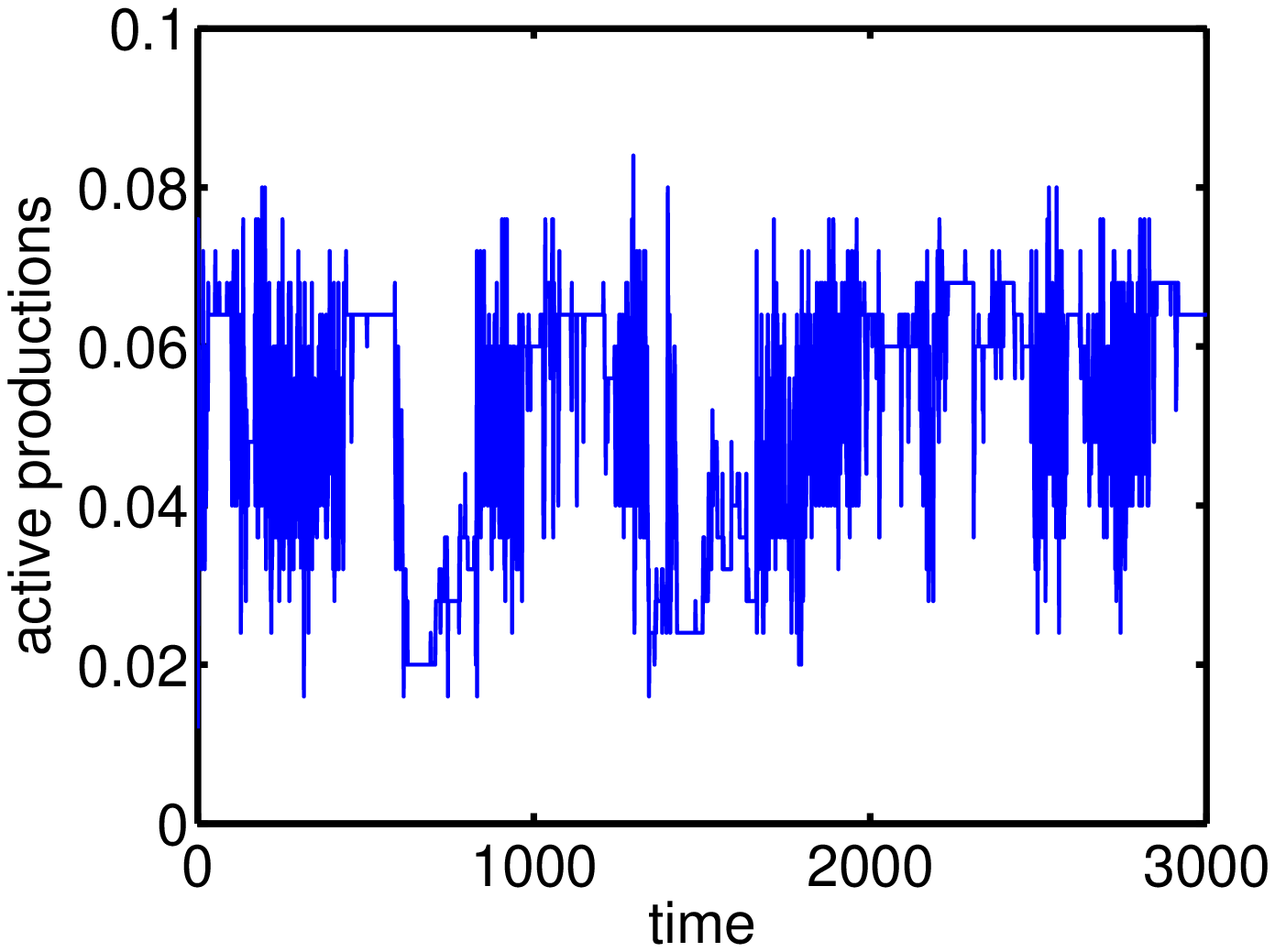}
\includegraphics[width=5.0cm]{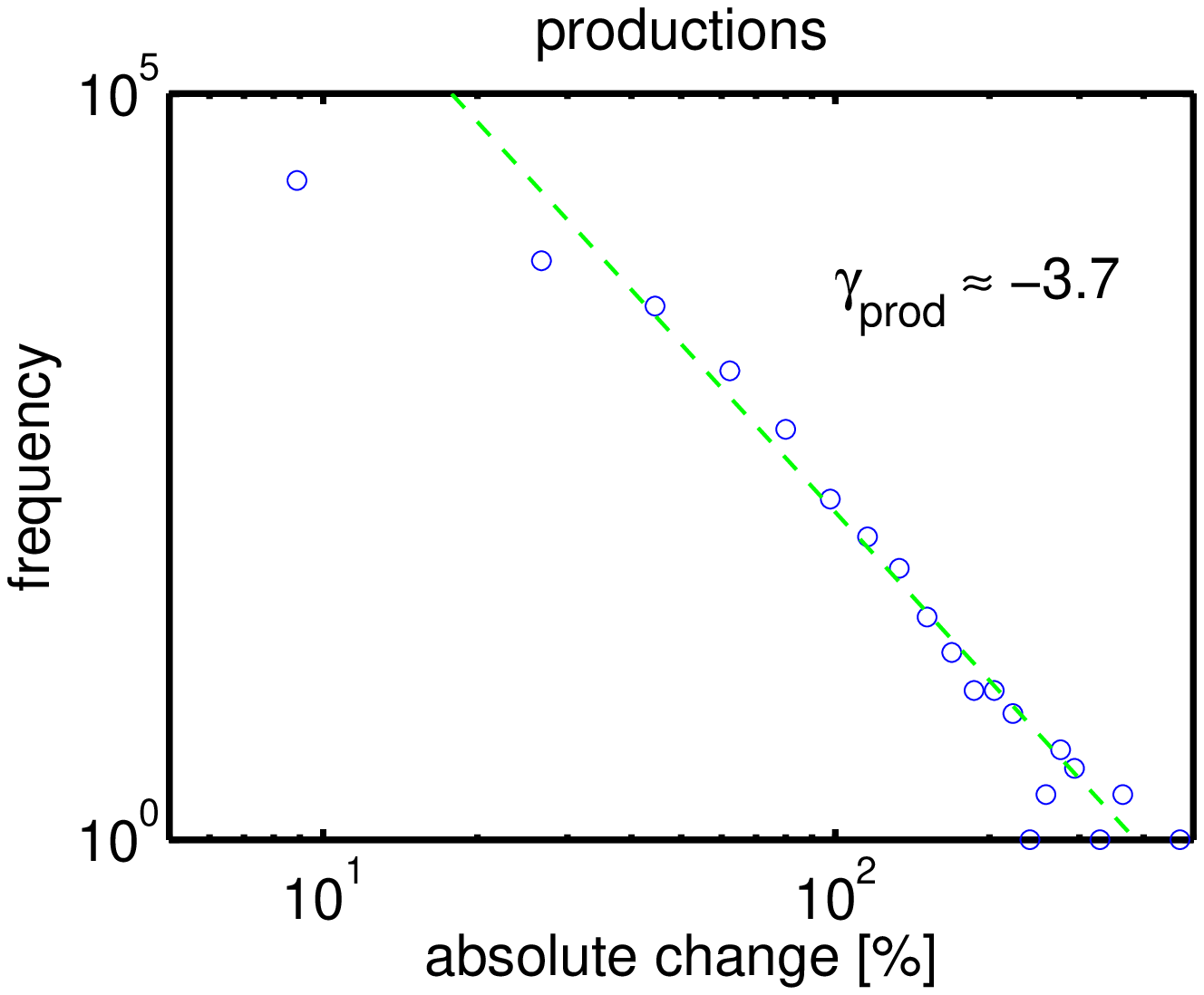}
\end{center}
\caption{Trajectories (a), number of total active productions at time $t$ (b), and (c) histogram of percent increments
of (b), for the more realistic competition process described in \ref{alternat}. Parameters $r^+=2.5$, $r^-=2$ and $p=10^{-3}$.
}  
\label{pro_dest}
\end{figure}

Technically, say the nonzero entries in $\alpha^+$ producing $l$ are $\alpha^+_{\mathbb i,l}=\alpha^+_{\mathbb j,l}=\alpha^+_{\mathbb k,l}=1$. 
We then impose the domination network for product $l$ in the following way: 
Suppose $\mathbb i = \{i_1, i_2\}$ and $\mathbb k = \{k_1, k_2\}$ then we encode the 
hierarchical suppression in $\alpha^-$ by setting $\alpha^-_{\mathbb i,k_1}=\alpha^-_{\mathbb i,k_2}=1$ and 
$\alpha^-_{\mathbb j,k_1}=\alpha^-_{\mathbb j,k_2}=\alpha^-_{\mathbb j,i_1}=\alpha^-_{\mathbb j,i_2}=1$. 
We repeat this step for each node $l \in \{1, \dots, N\}$. Except  for this alternative construction of $\alpha^-$ the model stays completely the same as before.
In Fig. \ref{pro_dest} we show the dynamical features of this model variant. Note that the slope in the production increments is somewhat 
steeper than for the original model. 

\subsection{Topology of production and destruction networks}

Results are surprisingly  robust with respect to changes in production/destruction topology $\alpha^{\pm}$. 
We have studied the dynamics on a {\em scale-free} production/destruction networks. 
In particular we investigated three possibilities to introduce power-law degree distributions: 
(i)  the number of productive/destructive sets per node (`in-degree') follows a power-law, 
(ii) the number occurrences of a given node in productive/destructive sets follows a power-law (`out-degree') or 
(iii) both in-degree' and `out-degree' are  distributed in a scale-free manner. 

For all choices the dynamical pattern of stochastic transitions between static and 
chaotic phases is retained. The presence of hubs somewhat stabilizes the system, i.e. 
generally with increasing exponent in the topological power-laws, lifetimes of plateaus increase.

Note, that the size of the productive/destructive sets is not limited to two. Exactly the same qualitative dynamical features of 
the system  are recovered for any values of productive/destructive set sizes, $n^{\pm}$, as long as 
$n^{+}>1$.
For $n^{\pm}=1$ the dynamics becomes linear and the system does not behave critical any more. 
In the same manner, the number of produced goods per productive set need not be restricted to one. 
One can think of one set producing/destroying more than one good. 
We have studies  variations in set sizes of this kind and found no changes in the  
qualitative dynamical behavior.

\subsection{Asymmetry in production and destruction}

For some circumstances it might be a realistic assumption that destructive influences have a 
larger impact than productive ones, i.e. that it is easier to destroy something than to `create' something. 
A simple way to study this is to enforce to set  $\Delta^{\pm}_i (t)$  to a negative value, as soon 
as there is at least one active destructive interaction pointing to good $i$, e.g. $N_i^{\rm destr}(t)>0 \to \Delta^{\pm}_i (t)=-1$.  
Note that this is similar to changing a majority rule to an unanimity rule \cite{opinion}.
The dynamical patterns we observe are robust concerning this variant. 
The same applies if we choose an `intermediate' model variant where we introduce 
a threshold $m>0$,  below which  $ \Delta^{\pm}_i (t)$ is set to be negative, i.e. 
if $ \Delta^{\pm}_i (t) < m \to  \Delta^{\pm}_i (t+1)=-1$.

\subsection{Modular structure of production/destruction networks}

It may be reasonable to impose a modular structure on the production/destruction topology.  
We constructed modules with different and random topologies which differ from each other by  different  values of 
$r^{\pm}$. Several modules (up to ten) of different densities  $r^{\pm}$ were then linked by a few 
connecting links. By increasing  the density of these connections between the modules, 
the system gradually undergoes a transition from a regime where each module behaves 
independently to a regime where the entire system's behavior becomes dominated by the most densely connected modules.

\subsection{Finite lifetime of goods and services}

It might be reasonable to consider that goods do not exist infinitely long, even in the case where no
destructive set points at it. One might want to introduce a decay rate of goods,  i.e. 
a good  $i$ decays with probability $\lambda$. 
For this model variant we find the same qualitative results as reported below. 
This decay rate can serve as a stochastic driving force, keeping the system from a 
frozen state. This means that even for $p=0$, for finite $\lambda$ the system 
does not freeze.

\subsection{Bounded rationality}

If a product {\em can} get produced it does not mean that it actually {\em will} get produced. 
To incorporate this possibility we say that if a good can get produced, it will  actually get produced with a probability $q$. 
This means that that if a good should be produced or destroyed 
according to Eq. (\ref{rule}), this happens only with probability $q$. With probability $1-q$ the opposite 
happens, i.e. if something should get produced -- it will not,  if something should be destroyed -- it will continue to exist. 
This probability can either be seen as a spontaneous idea 
of an entrepreneur or as a lack of rationality.
This variant is formally almost exactly the same as driving the system with the innovation parameter $p$;  
the scenarios only differ if the sign of $\Delta^{\pm} (t)$ changes exactly when the $p$ or $q$ event happens, which is rare.

\begin{figure}[t]
\begin{center}
\includegraphics[height=5.8cm]{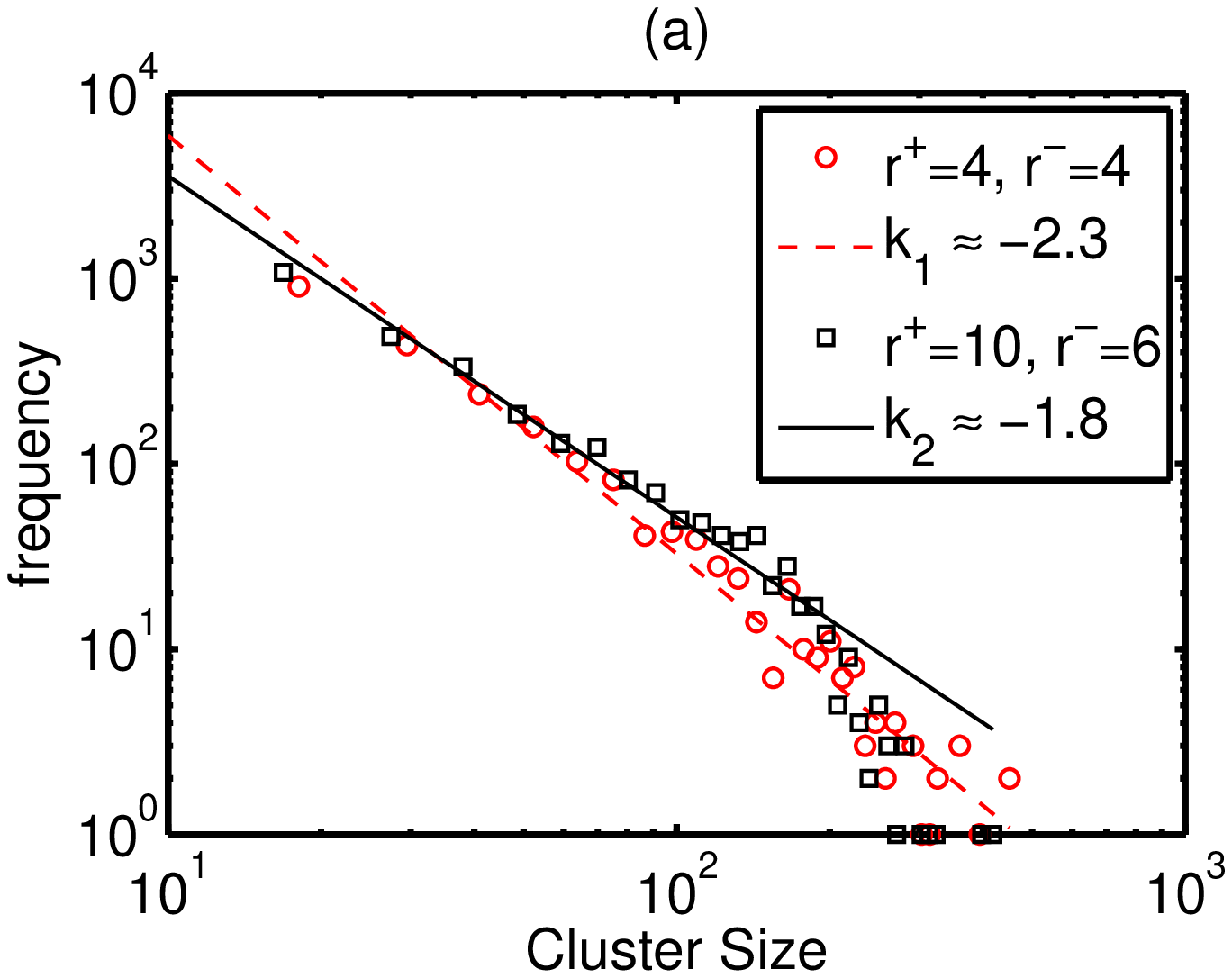}
\includegraphics[height=5.8cm]{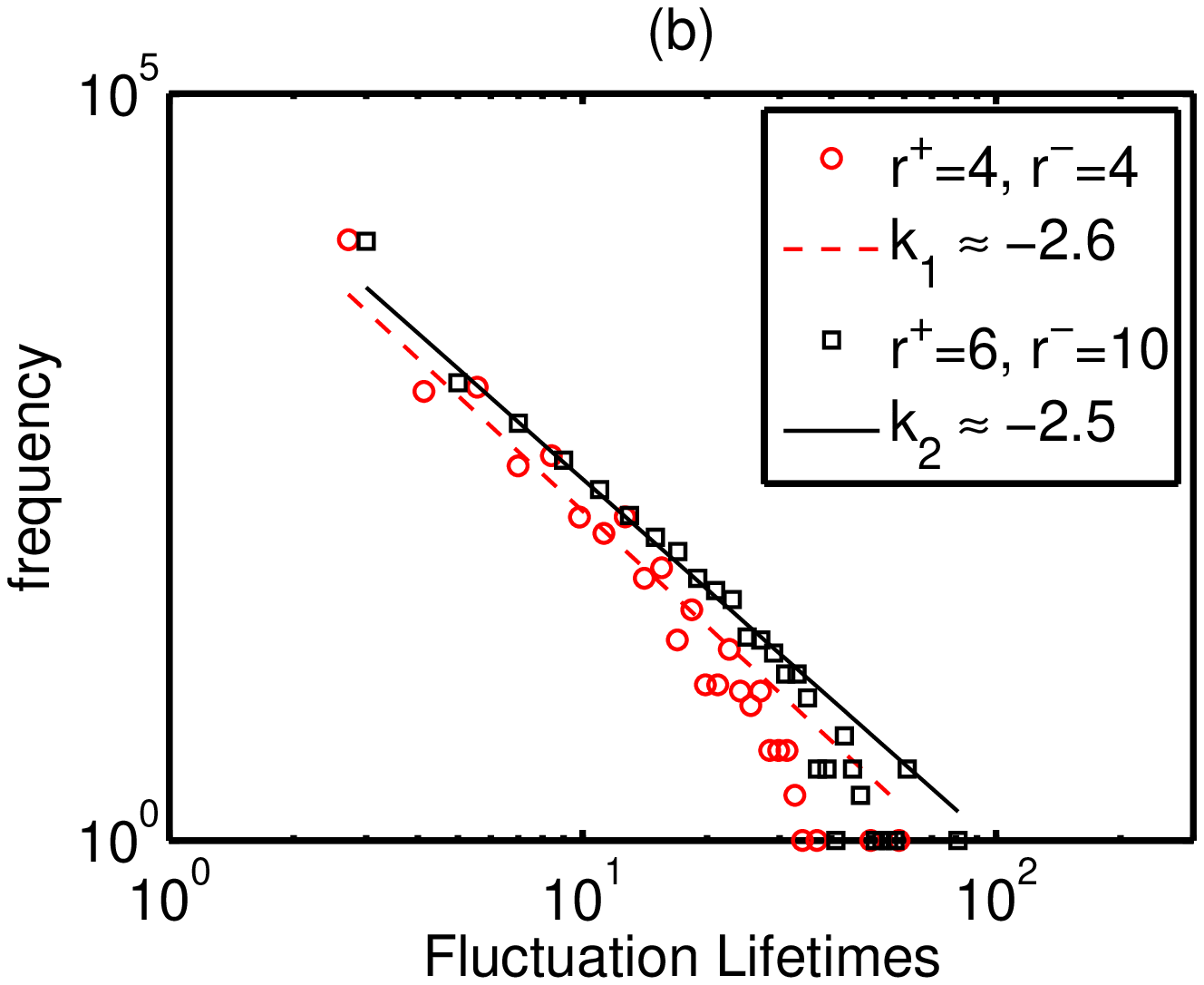}
\end{center}
\caption{
Cluster size distribution (a) and fluctuation lifetime distribution (b) of a system of size $N=1000$ for different densities $r^+$ and $r^-$.
The slopes for power-law fits for $r^+=r^-=4$ are $k_1=-2.3$ for the sizes, and $k_1=-2.6$ for fluctuation durations, 
while for $r^+=10$ and $r^-=6$,  we get $k_2=-1.8$ for sizes, and $k_2=-2.5$ for durations. Given that these 
parameter settings correspond to highly different scenarios, the similarity of power exponents 
indicates some degree of robustness.  
}
\label{ClustSizes}
\end{figure}

\subsection{Variations in the update}

The qualitative behavior of the model does not change if we employ a parallel update or a 
sequential update in a deterministic order. The only impact of these changes is 
that the system needs longer to find frozen states in parallel updating.

\section{Understanding Schumpeterian dynamics}

The high level of generality of the presented model of Schumpeterian dynamics allows several ways of understanding. 
We discuss two ways. First we show a  correspondence to self-organized critical (SOC) sandpile models.
Second, we understand Schumpeterian dynamics on the basis of the eigenvalues of the active production network.

\subsection{Schumpeterian dynamics is a SOC sandpile}

Schumpeterian dynamics in our implementation  can be seen as a self-organized critical system. To see the direct 
similarities to  a sandpile model \cite{bak87} we proceed in the following way: 
set $p=0$ and wait until the system reached a frozen state, which we define as one or less changes in $\vec \sigma$ 
occurring over five iterations \footnote{More precisely, we count the number of changes 
of states between two consecutive time steps, i.e. the
quantity $\delta \sigma (t) = \sum_i | \sigma_i(t+1) - \sigma_i(t) |$.  If  $\delta \sigma (t)$ is not
higher than one, within five iterations, i.e. $\delta \sigma (t') \leq 1$ $\forall$ $t' \in \{t, t+1, \dots, t+5\}$,
we call the state of the system quasi stationary. Now if the system is in such a state and gets perturbed,
our definition ensures that we still remain in the same stationary state if this perturbation dies out within
five iterations. On the other hand, if the perturbation triggers a cascade and spreads over the system,
there will soon be more than one update between two iteration and the system escapes the stationary state.}.
We then flip one randomly chosen component $\sigma_i$. This perturbation may or may not trigger successive updates. 
In Fig. \ref{ClustSizes} (a) the cluster size (total number of goods that get updated as a consequence 
of this perturbation) distribution is shown . The observed power-laws reflect typical features of self-organized criticality: 
one spontaneous (irrational) event may trigger an  avalanche of restructuring in the economy;  the power-laws demonstrate  that 
large events of macroscopic size are by no means rare events, but rather the rule than the exception. 
We show the distribution of `fluctuation-lifetimes' in Fig. \ref{ClustSizes} (b), i.e. 
the number of iterations which the system needs  to arrive at a frozen state. The distribution of `lifetimes'  also 
follows a power-law, which  confirms the existence of self-organized criticality in our model in the sense of \cite{bak87}.

\subsection{Eigenvalues, keystone productions}

\begin{figure}[t]
\begin{center}
\includegraphics[height=3.8cm]{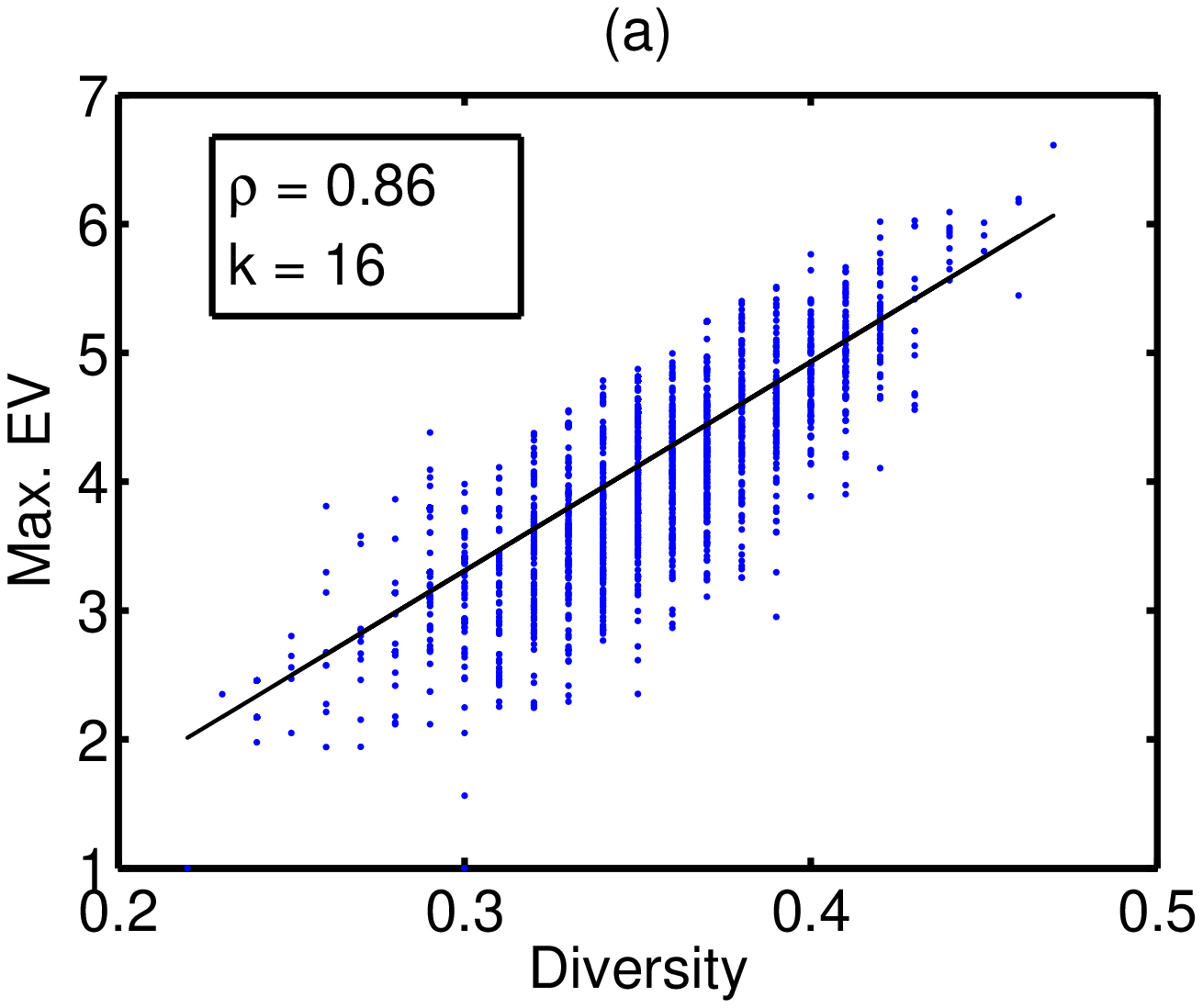} 
\includegraphics[height=3.8cm]{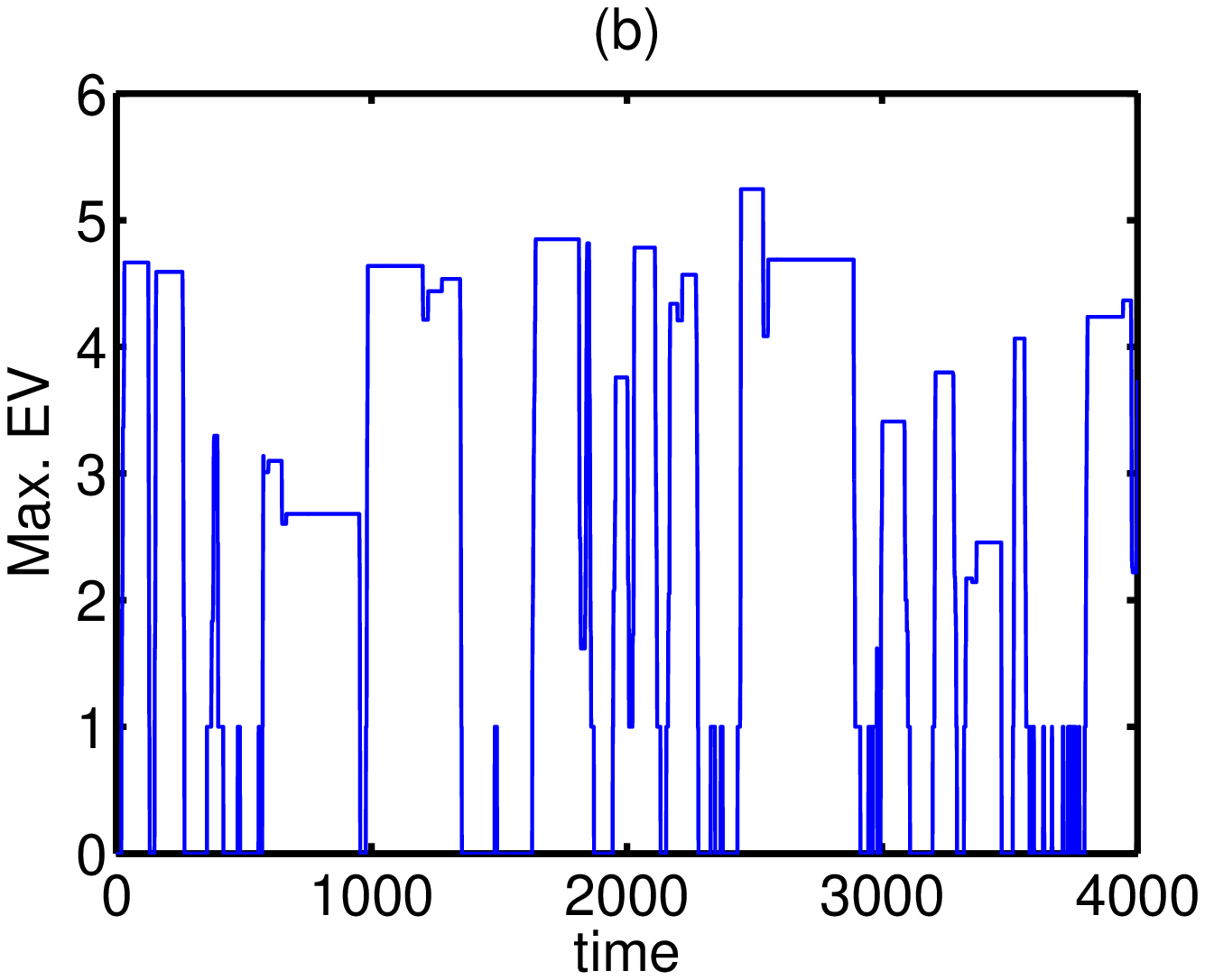}
\includegraphics[height=3.8cm]{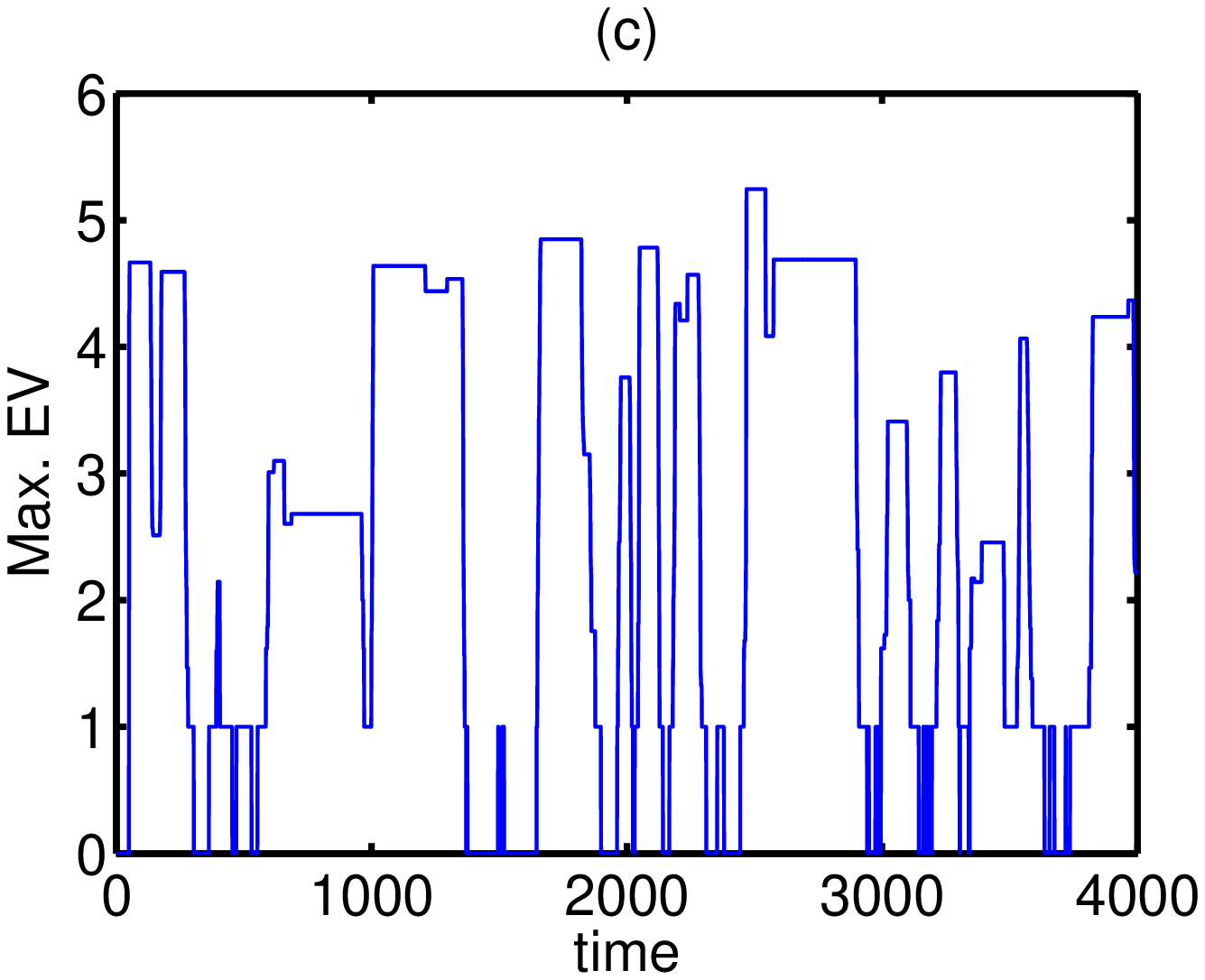}
\end{center}
\caption{
(a) At each time step we construct the active production network $A(t)$ for the run of Fig. \ref{trajectory}, with $h=1$ and $T=1$. 
Its maximal eigenvalue is plotted vs. the system's diversity at that time $t$. Every point represents one  timepoint.
(b)-(c) Comparison of the maximum EV when computed with filtering, using $T=20$, $h=0.95$ (b) and  $T=50$, $h=0.8$ (c)
 for the  same run shown in  Fig. \ref{trajectory}.
 }
 \label{MaxEV}
\end{figure}

To understand Schumpeterian dynamics it seems natural to study  the topology of the active production network. In particular 
the question arises of how topology is related to the outcome of the dynamical system. 
The simplest  quantitative measure related to dynamics on networks is to compute the 
maximum real eigenvalue of the active production network $A(t)$. 
In Fig. \ref{MaxEV}  (a)
we show a plot of the diversity vs. the maximum real eigenvalue of the adjacency  associated to the 
active production  adjacency network.
The latter have been constructed as described in Section \ref{apn}, however  without using the filtering, i.e. $h=1$ and $T=1$.
There is a correlation of about $\rho\sim 0.85$ the slope is $\sim16$.

In Figs. \ref{MaxEV} (b) and (c) we demonstrate that the dependence of the maximum eigenvalue is rather independent of the choice 
$h$ and $T$, and thus justifies the approach, whenever  $h$ and $T$ remain in reasonable limits.  
For the chaotic phases the maximal eigenvalue is mostly zero which indicates an directed acyclic graph. 
When  a maximal eigenvalue of one is found this indicates one or more simple cycles \cite{farmer}.  
On the plateaus we typically find values larger than one which is a sign of a larger number of interconnected cycles in $A$. 
In this sense the plateaus are characterized by a long-lasting high level of 'cooperation', 
whereas in the chaotic phase cooperation between cyclically driven production paths is absent. 
This is the topological manifestation of collective organization which the model produces.

\begin{sidewaysfigure}
\centering
\vspace{13cm}
\scalebox{1.15}{\includegraphics[height=4.0cm]{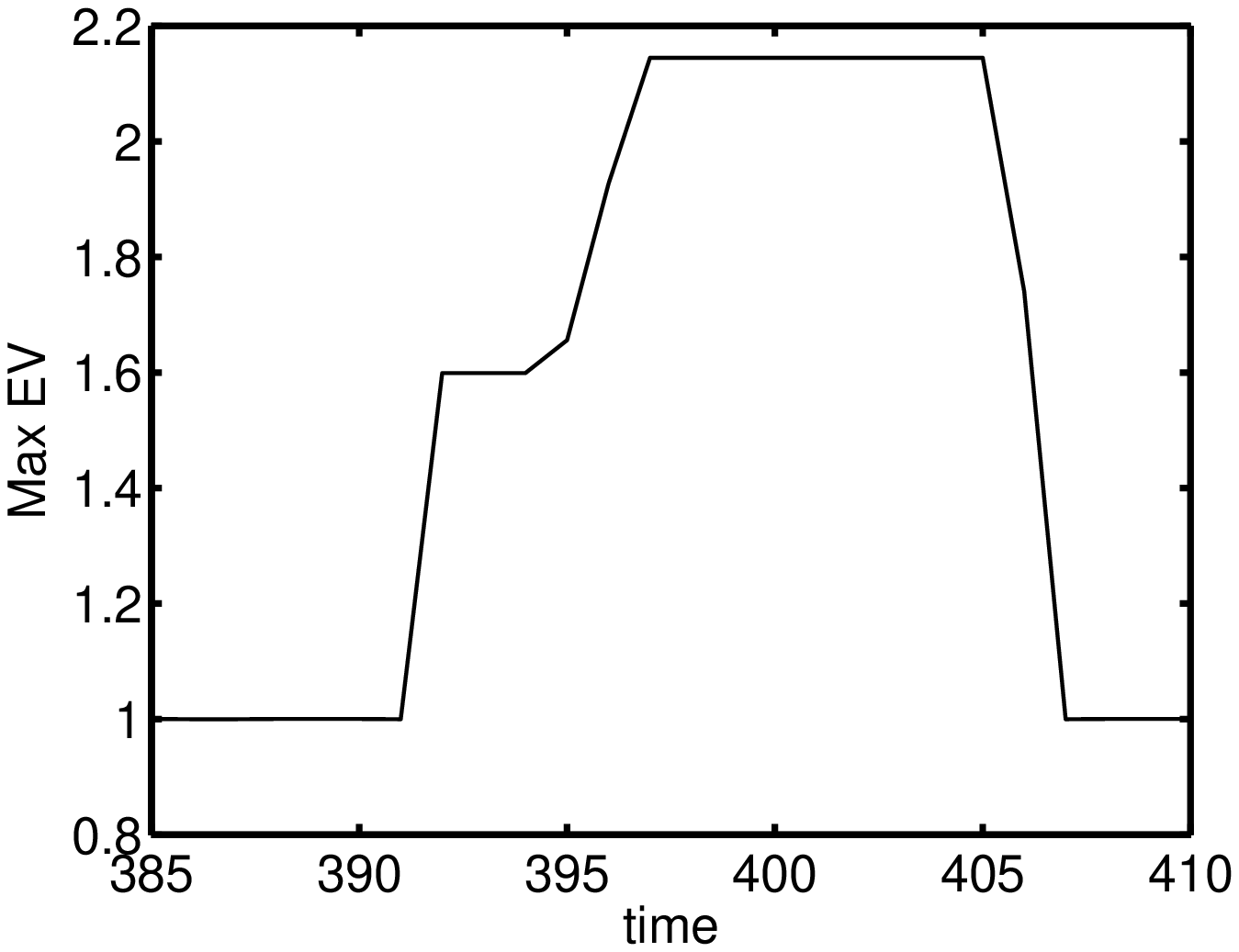}\includegraphics[height=4.0cm]{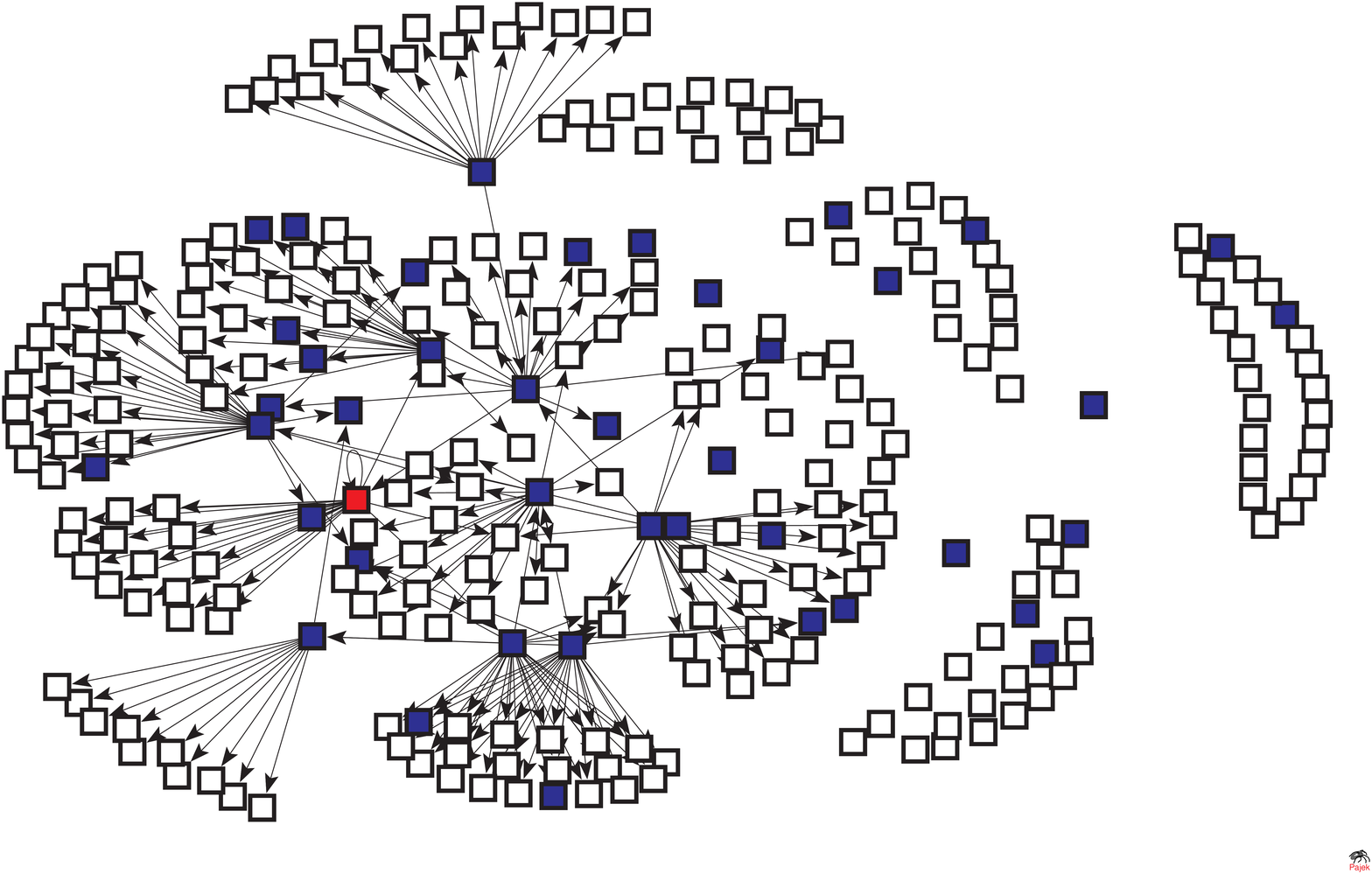}\hspace{-1.2cm}\mbox{t=391}\includegraphics[height=4.0cm]{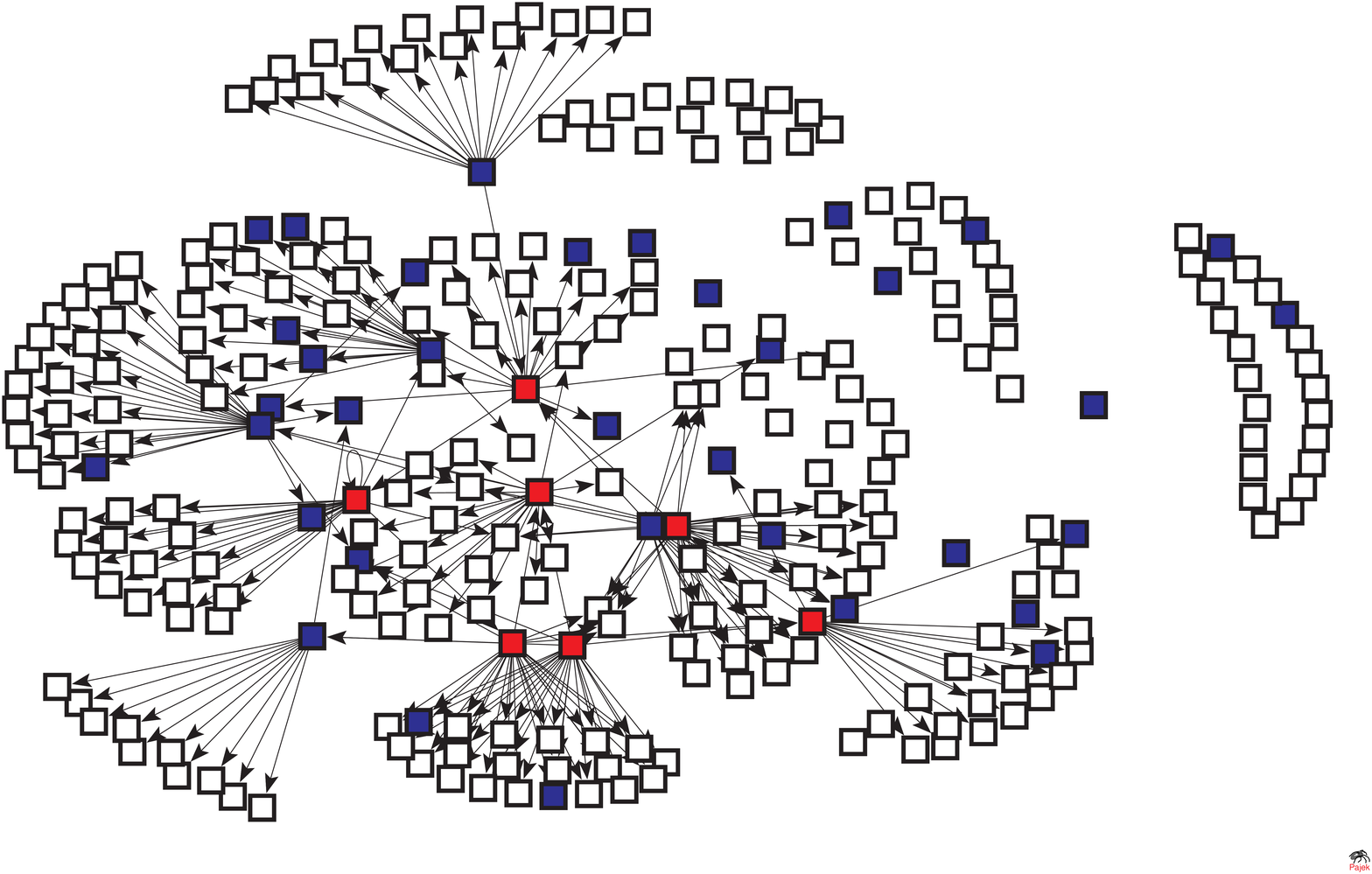}\hspace{-1.2cm}\mbox{t=392}}
\scalebox{1.15}{\includegraphics[height=4.0cm]{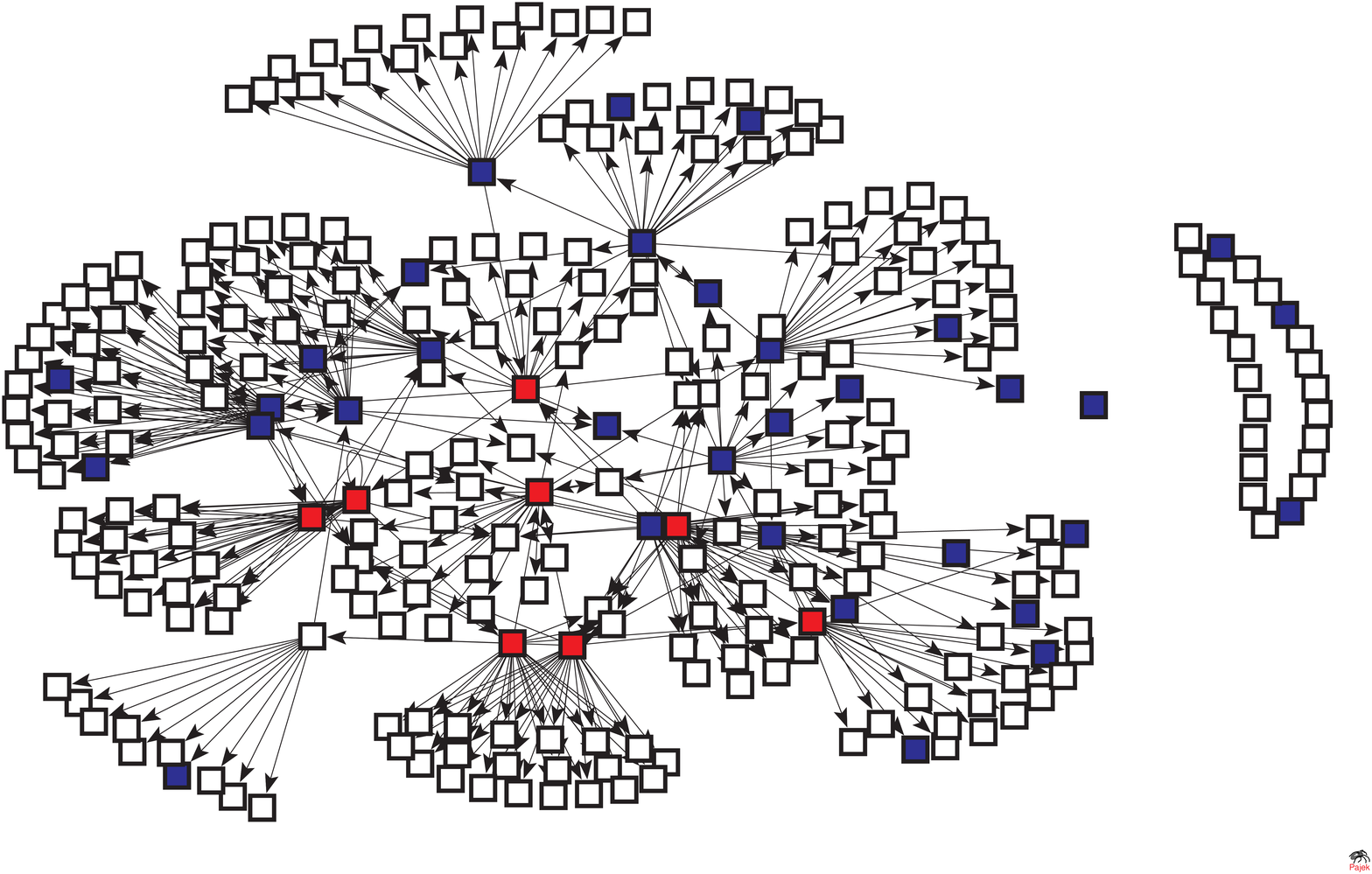}\hspace{-1.2cm}\mbox{t=396}\includegraphics[height=4.0cm]{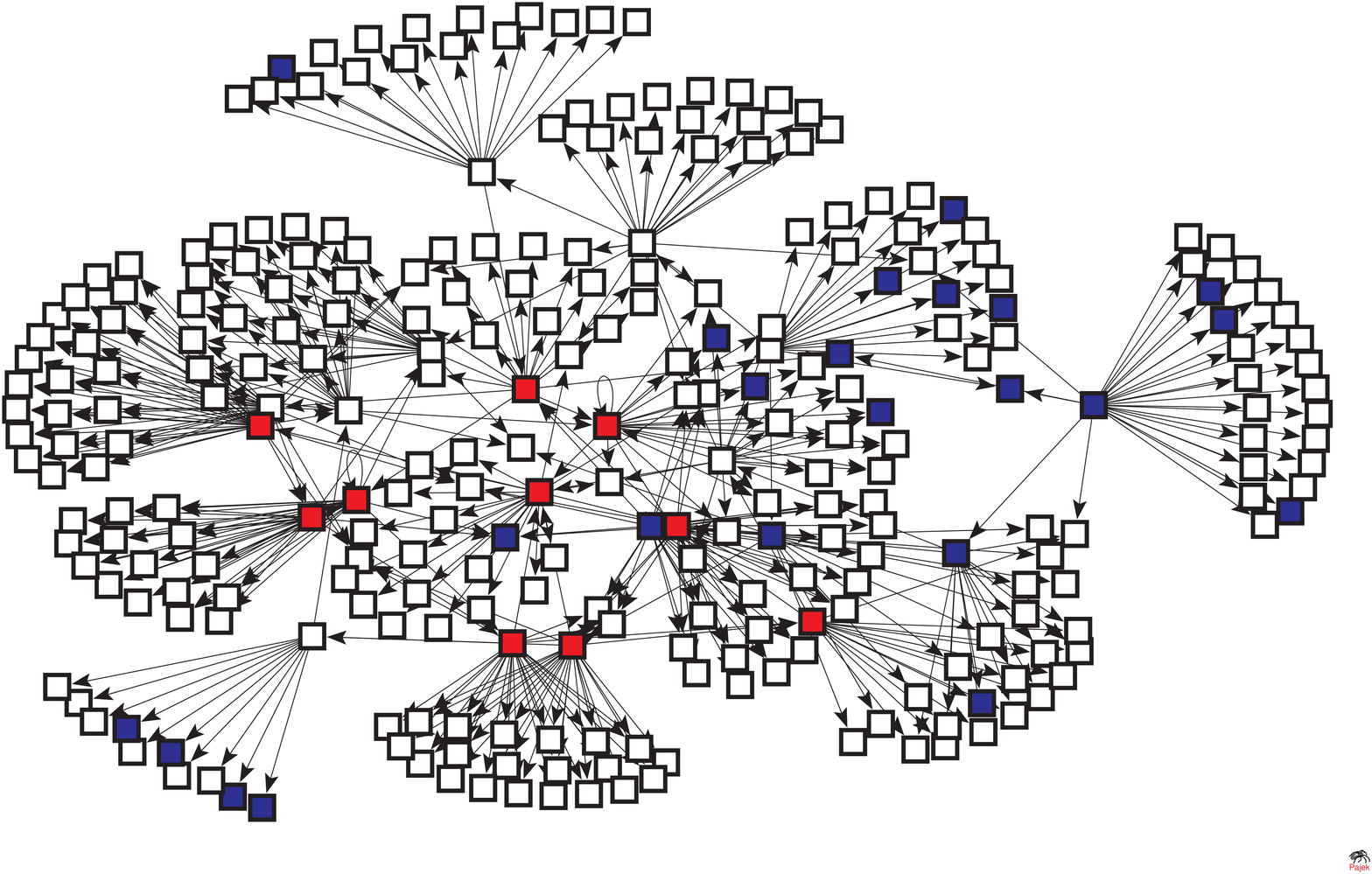}\hspace{-1.2cm}\mbox{t=400}\includegraphics[height=4.0cm]{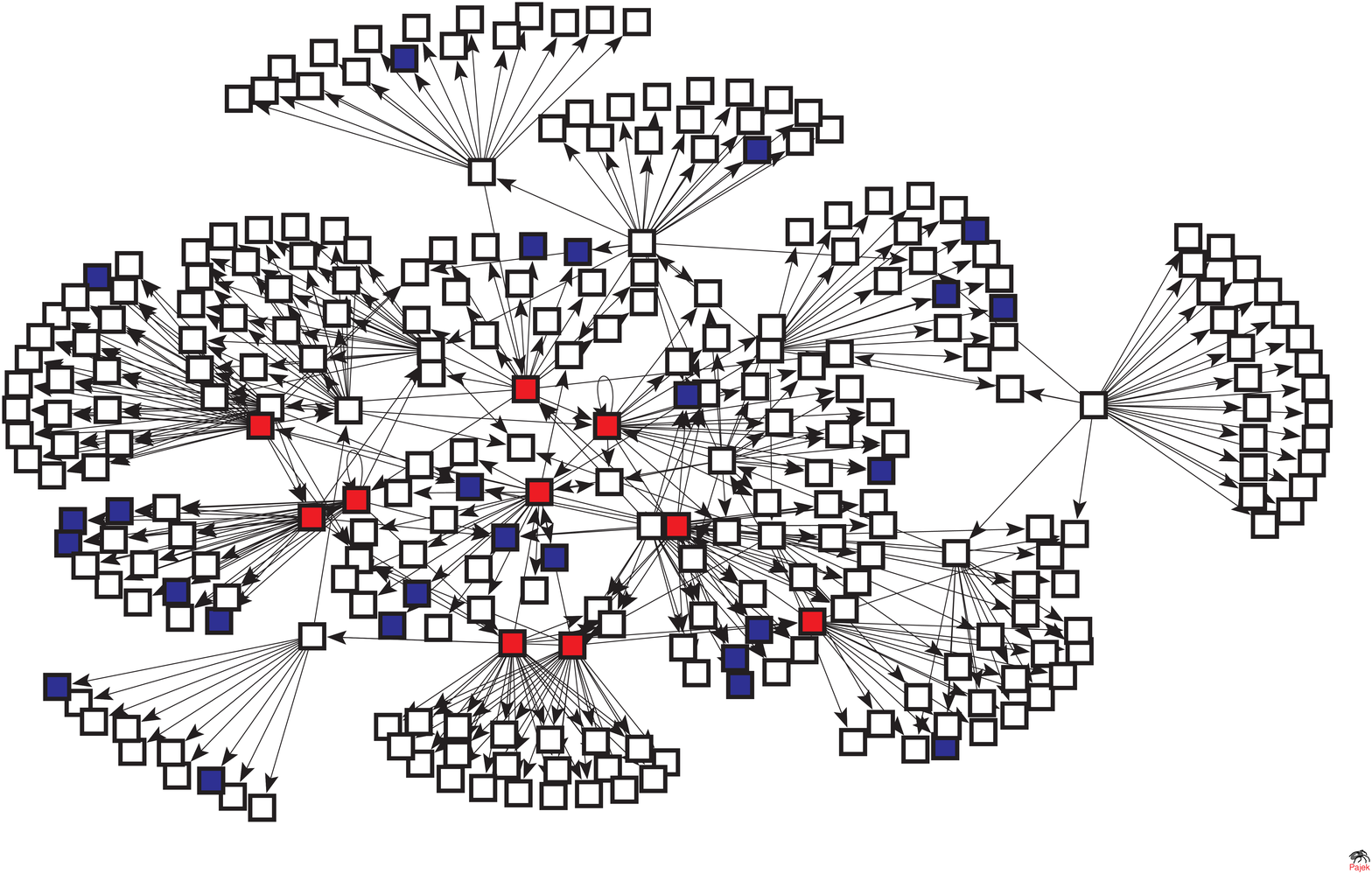}\hspace{-1.2cm}\mbox{t=405}} 
\scalebox{1.15}{\includegraphics[height=4.0cm]{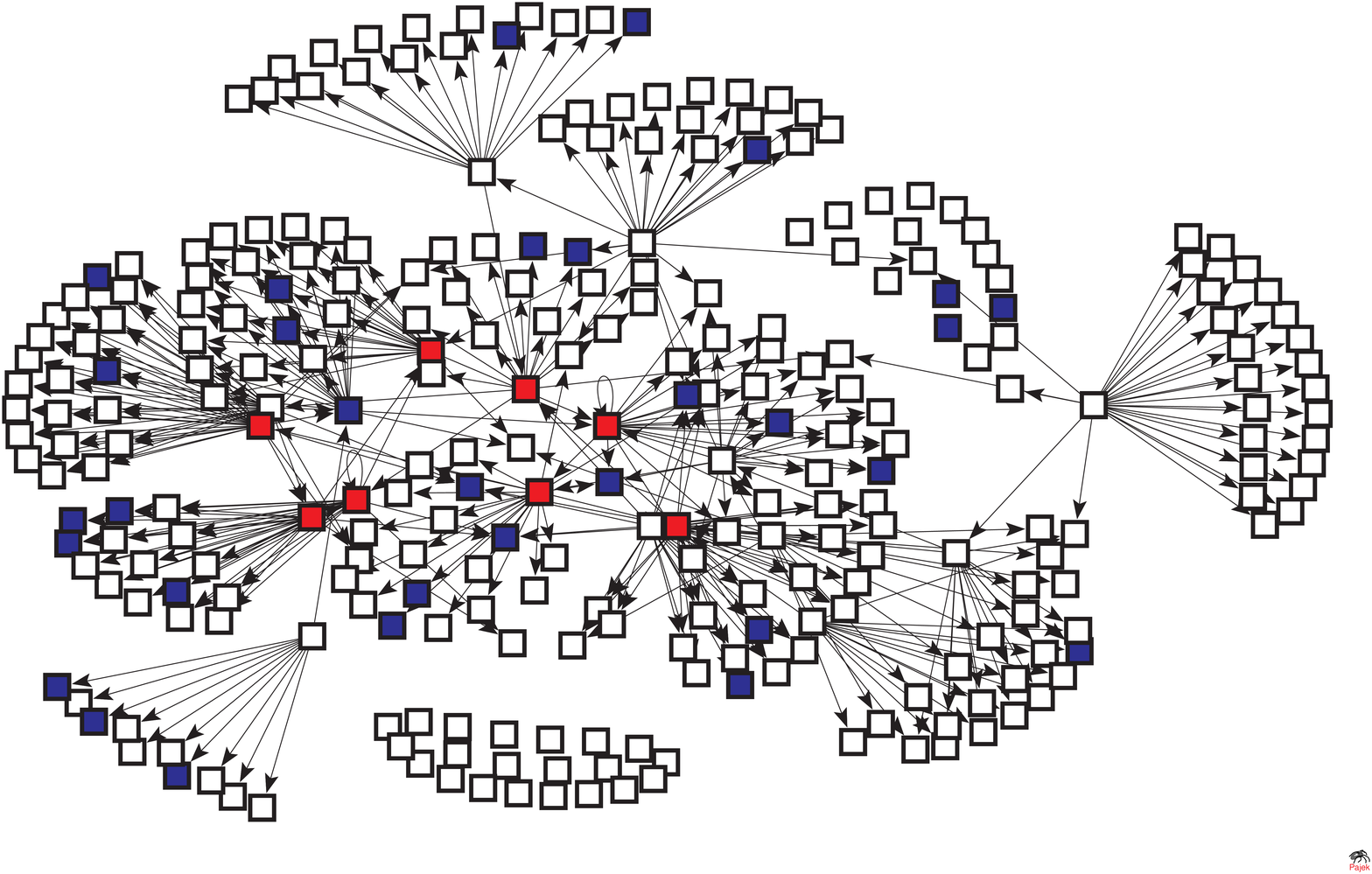}\hspace{-1.2cm}\mbox{t=406}\includegraphics[height=4.0cm]{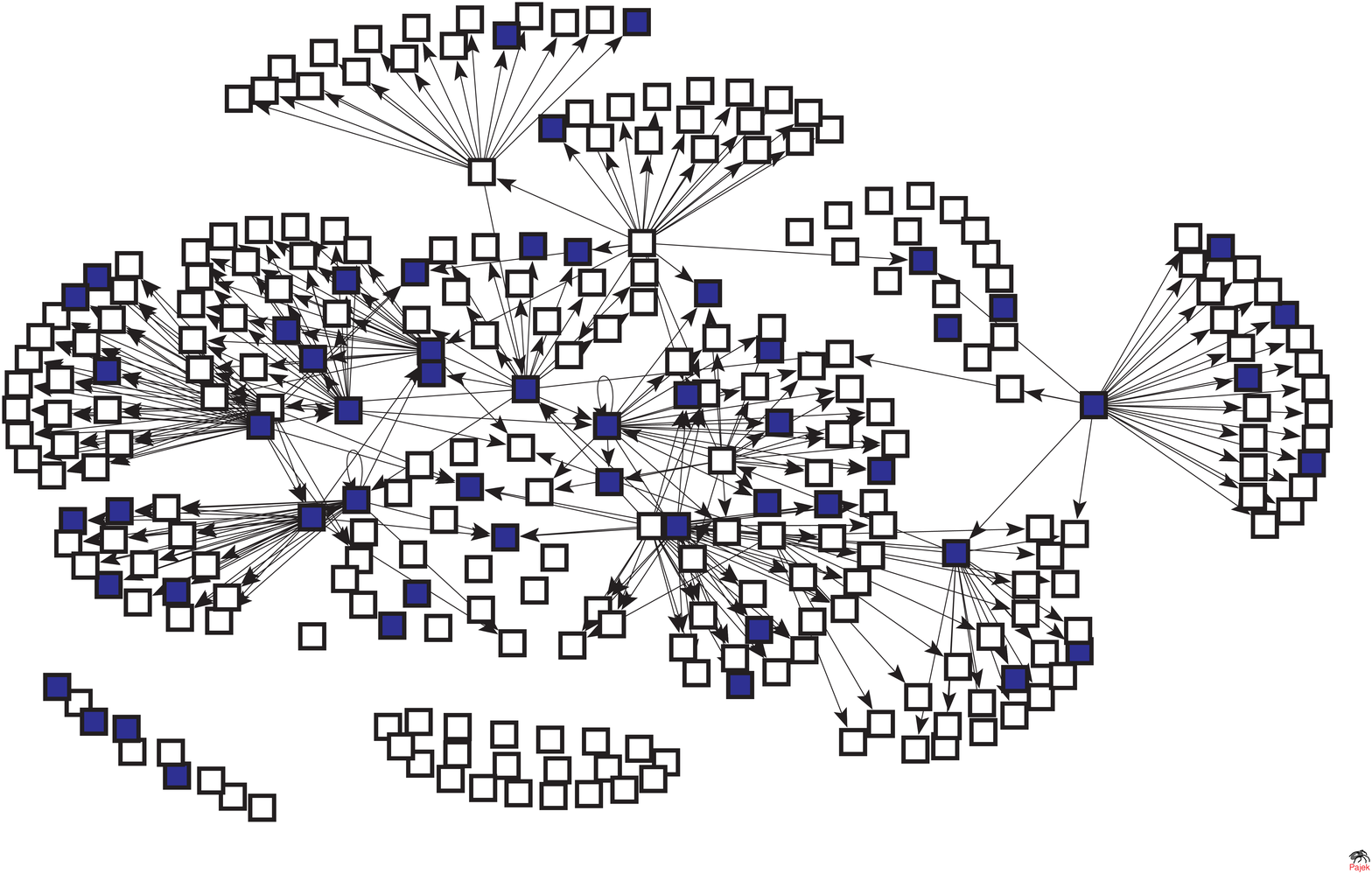}\hspace{-1.2cm}\mbox{t=407}\includegraphics[height=4.0cm]{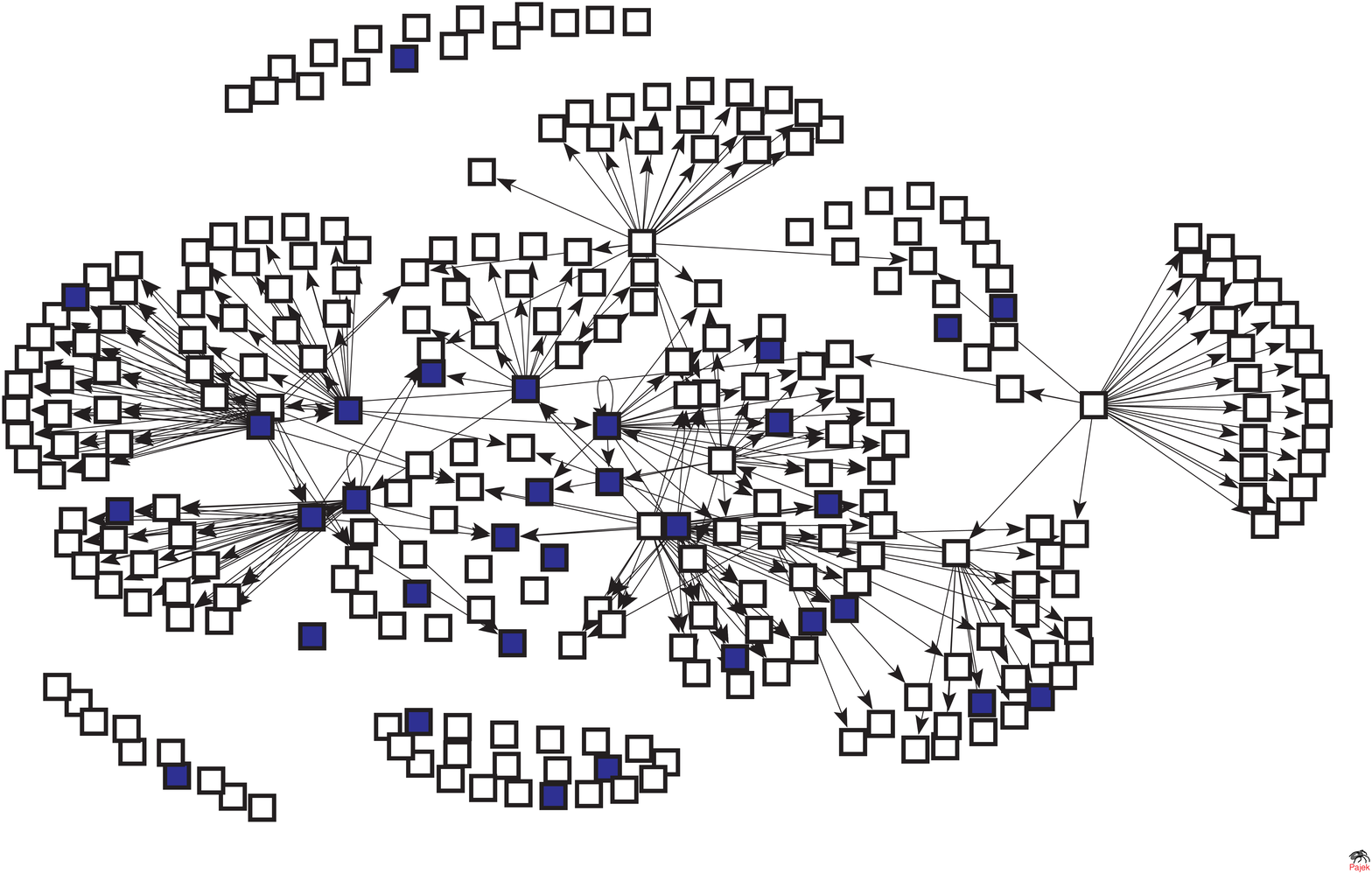}\hspace{-1.2cm}\mbox{t=408}} 
\caption{Onset and breakdown of cooperation. We show the trajectory of the maximum eigenvalue of $A$ in (a) for $T=50$ and $h=0.8$.  
In the particular example the eigenvalue starts at around 1 builds up to a plateau at about 2.1 and then drops to about 1 again. 
The effective production networks  along this trajectory are shown. At first the network contains simple cycles (maximal eigenvalue equals one). 
At each time step we show 'keystone' productions as red nodes. A keystone production is defined 
as a node which -- if it was removed from $A(t)$ -- would result in a reduction of the maximum eigenvalue 
of more than $10\%$ (when compared to the maximum eigenvalue of $A$).Typically keystone nodes are components of cycles. 
Note, active destruction graphs can be constructed in the same way but will always be much sparser that active production networks.
}
\label{nets}
\end{sidewaysfigure}

The situation here is similar to what was found in a model of biological evolution  by Jain and Krishna \cite{jainkrishna}. 
Unlike the dynamics of the active production/destruction network $A$,  in their model, Jain and Krishna update their 
interaction matrix through a selection mechanism. They were among the first 
\footnote{In \cite{farmer} this has been already anticipated before.}
to relate the topology of the dynamical interaction matrix to the diversity of the system. 
In particular they could show 
that the highly populated phases in the system is always associated with autocatalytic cycles and keystone species, i.e. species building up these cycles. 
Drastic increase in diversity is associated to the spontaneous  formation of such cycles, the decline 
of species diversity is triggered by breaking a cycle. 
Even though we do not have any explicit selection mechanism in the model the relation between topological structure of $A$
and and the state of the dynamical system is the same as in \cite{jainkrishna}.
To explicitly show the relation between eigenvalues, cycles and product diversity we show the trajectory  of the maximum 
eigenvalue of $A$  
together with snapshots of active production networks along the trajectory in Fig.  \ref{nets}.

\section{Discussion}

We try to capture the essence of Schumpeterian economic dynamics in a simple dynamical 
model, where goods and services co-evolve with their `activated' production networks. 
New goods and services are endogenously produced through re-combinations of existing goods. 
New goods entering the market may compete against already existing goods and as a result of this competition mechanism 
existing goods may be driven out from the market - often causing cascades of secondary defects (Schumpeterian 
gales of destruction). 

The model leads to a generic dynamics characterized by phases of relative economic stability followed by phases 
of massive restructuring of markets -- which can be interpreted as Schumpeterian  business `cycles'. 
Cascades of construction and destruction produce typical power-law distributions,  
both in cascade size and in times of restructuring periods. The associated power exponents are rather 
robust under the chosen parameters governing the density of productive/destructive rules. 
The model can be fully understood as a self-organized critical system. 

Alternatively the diversity dynamics generated by the model can be understood along the same lines as the 
evolution model of Jain and Krishna \cite{jainkrishna}.  
As in \cite{jainkrishna} we are able to relate the diversity of the system to topological properties of the 
active production network. The maximum real eigenvalue of the active production networks correlates strongly with diversity. 
Further we were able to identify `keystone'  productions, which -- when removed -- dramatically reduce the maximum eigenvalue.
To a certain extend it is also possible to relate the model to the Bak-Sneppen model \cite{baksneppen}, in the sense that 
the local  quantity $\Delta_i^{\pm}$ can be thought of playing the role of    the random fitness in the  Bak-Sneppen model. 

Model timeseries of product diversity and productivity  reproduce several stylized facts of economic timeseries on long 
timescales such as GDP or business failures, including non-Gaussian fat tailed distributions, volatility clustering etc. 
We have studied a series of more realistic model variants. Remarkably, 
the majority of the statistical results  holds qualitatively also for these variants, and a certain degree 
of universality of the model is indicated. So far we have not analyzed universality issues in much detail. 

Finally we mention that parts of the model are exactly solvable. The problem for a constructive dynamics only has been 
solved analytically in the long time limit in \cite{htk1}, showing that such systems exists in two 
phases. A low diversity phase is separated from a high diversity phase through a first order phase transition.
By superimposing the destructive dynamics here in this model, provides a mechanism to shift from one regime to the 
other. However, this superposition makes it impossible to solve the model analytically, 
without recourse to non-Gaussian  spinglass techniques.   
\\
 
This work was supported by Austrian Science Fund FWF P 19132. \\

\end{document}